\documentclass[aps,showpacs,amsmath,amssymb,prc,nofootinbib]{revtex4}
\usepackage{graphicx}

\begin{document}


\title{$J/\psi$ dissociation by light mesons in an extended Nambu Jona-Lasinio model}

\author{A.~Bourque}
\author{C.~Gale}
\affiliation{Department of Physics, McGill University \\ 3660 University Street, Montreal, QC,
Canada H3A 2T8}


\date{\today}

\begin{abstract}
A model for the dissociation of the $J/\psi$ is proposed, where 
chiral symmetry is properly implemented. Abnormal parity interactions and mesonic form factors naturally arise
from the underlying quark sub-structure. Analytic confinement of the light quarks is obtained through an appropriate
choice of quark interaction kernels. Dissociation cross sections of the $J/\psi$ by either a $\pi$ or a $\rho$ meson 
are then evaluated and discussed.\end{abstract}


\maketitle

\section{Introduction}
Lattice simulations of quantum chromodynamics (QCD) predict a transition from hadronic matter to a plasma of quarks
and gluons at a critical temperature of $T_c = 175 \pm 10$ MeV \cite{Pet05,Sat05}. To reproduce such a condition in
a terrestrial environment, heavy ions are collided at relativistic energies. A typical space-time evolution involves the collision system 
going through various phases including possibly the elusive quark-gluon plasma (QGP). To find an unambiguous
signature of this new state of matter proves challenging as other stages of the fireball expansion
can also make contributions, which will thus constitute background.

One popular probe of the quark-gluon plasma is the charmonium yield modification first suggested 
by Matsui and Satz in a seminal paper \cite{Mat86}. In their original scenario, the charmonia produced in the earliest stage of
the collision is expected to be suppressed by the QGP due to color screening \cite{Kha97}. But late stage
hadronic dissociation could generally also
occur, thus making the sources of the suppression difficult to disentangle: understanding the charmonium
dissociation within a hadronic gas becomes essential. However, very little is known  experimentally about 
these hadronic dissociation processes and one has to rely on theoretical calculations.

Most studies focus on dissociation channels of the charmonia by pions as it is the most abundant particle in
the produced hadronic gas. Moreover, since, in a thermal gas at realtistic temperature, the pions have just enough energy to dissociate
the charmonium ground-state, the $J/\psi$, the cross section near threshold is of particular interest.
Various approaches can be used including non-relativistic potential models
\cite{Bar03,Mar95,Won00,Won02}, QCD sum rules \cite{Dur03,Nav00,Bra01,Nav02,Nav02_2,Mat02,Dur03_2,Aze04}, and
constituent-quark based formalisms \cite{Mai05,Pol00,Bla01,Dea03,Mai04,Iva05,Lap06}. Alternatively, 
phenomenological Lagrangians can also be employed \cite{Mat98,Lin00,Lin00_2,Hag00,Hag01,Oh01}. However,  
 the implementation of chiral symmetry has not uniformly been done in all Lagrangian models \cite{Nav01}. This could then have an important phenomenological consequence, as it is expected to soften the
cross section near the production threshold.  

In Ref.~\cite{bou05_1, bou08}, the effect of chiral symmetry on the dissociation
cross sections was
investigated within a chiral symmetric phenomenological Lagrangian approach. It was shown that for a certain class of interactions the
dissociation transition amplitudes should vanish as the pion momentum goes to zero (soft-pion theorem). Although, a reduction was observed
near threshold, the effect of introducing the so-called abnormal parity interactions was at least as important as that of the implementation of chiral symmetry. This observation could be traced back 
to the fact that these abnormal parity interactions circumvent the soft-pion theorem.
The results of Ref.~\cite{bou08}, in particular the overall magnitude of the calculate cross sections, depended heavily on the parameterization of the ad-hoc form factors. It is the purpose of this article to address this issue by
proposing a chiral-symmetric model of the $J/\psi$--dissociation whereby the
form factors naturally arise from the underlying quark structure. This
constituent-quark framework builds on models presented in Refs.~\cite{Mai05,Pol00,Bla01,Dea03,Mai04,Iva05,Lap06}.
However, strickly speaking, the model presented here is an extension to the charm sector of Refs.~\cite{Bow95,Pla98,Pla02}
which itself is a generalisation of the Nambu Jona-Lasinio (NJL) model \cite{Nam61,Wei91}
where the four--point interaction kernels are non-local and chiral symmetry is implemented
at the quark level. Within this formalism, the mesonic form factors can
then be calculated, and  by choosing the appropriate light--quark kernels, 
it is possible to push to higher energies, or even remove, the 
unphysical production threshold of the $\rho$ meson into the continuum (i.e, into a $\bar q-q$ pair) which at
zero temperature and density should not occur. This then also permits the calculation of the
$\rho$-induced $J/\psi$ dissociation which is used to assess the effect of light resonances on the overall
dissociation strength.

This article is organized as follows: after introducing the quark interaction kernels and
considering their
properties, quark propagators are discussed with an emphasis on ways of generating quark confinement. Meson
bound states are then found along with meson-quark vertex functions. Three-- and four--point meson
interactions mediated through quark loops are written down. Finally, after fixing the various parameters using a combination 
of lattice results and empirical information, the behaviors of meson propagators, 
vertices, and cross sections are examined. The analytic continuation prescriptions are detailed in Appendix \ref{analytic}.
In Appendix~\ref{currents}, the isovector axial Ward identity is
explicitly checked, while in Appendix \ref{Decays} decay processes used to fix
the parameters are evaluated. 
Finally, the amplitudes of the various
dissociation channels studied here can be found in Appendix~\ref{amplitudes}.

\section{Quark interaction kernels}
In this model, chiral symmetry is implemented at the quark level. As in Ref.~\cite{bou08}, the  
mesonic content includes the $\pi$, $\rho$, $J/\psi$, 
$D$, $D^*$, and the chiral partners of the open charmed mesons, namely $D^*_0$ and $D_1$
mesons,  in order to describe the
dissociation processes of interest. Besides these, the $\sigma$ and $a_1$ mesons will 
also naturally appear in such an approach due to chiral symmetry. However, as for the NJL model, only
global color invariance is introduced allowing to account for 
the number of quark colors, $N_c$, in QCD. Moreover, only color singlets are
written down.

With this in mind, the minimal action is
\begin{eqnarray}
\mathcal{S} = &\int& dx \left \{\bar q(x)\left(i\displaystyle{\not}\partial -m^q_{c}\right)q(x) +
\bar Q(x)\left(i\displaystyle{\not}\partial -m^Q_{c}\right)Q(x)\right\} +S_{int}
\label{LENJL}
\end{eqnarray}
where $q$ and $Q$ are the fermion fields for the light and heavy quarks, respectively, and the interactions
are decomposed into
\begin{equation}
S_{int} = \mathcal{S}^{qq}_{int} + \mathcal{S}^{qQ}_{int}+\mathcal{S}^{QQ}_{int}
\end{equation}
with
\begin{equation}
\mathcal{S}^{f_1f_2}_{int} = \left[\prod_{k=1}^4\int dx_k \right]
K^{f_1f_2}_{abcd}(x_1,x_2,x_3,x_4) \bar
\psi^a_{f_1}(x_1)\psi^b_{f_1}(x_2) \bar \psi_{f_1}^c(x_3)\psi^d_{f_2}
(x_4).
\label{LENJL_interactions}
\end{equation}
The kernels, $K$, can be further decomposed into 
\begin{equation}
K^{f_1f_2}_{abcd}(x_1,x_2,x_3,x_4) = \sum_i H^{f_1f_2}_i(x_1,x_2,x_3,x_4) \left
(\bar \Gamma^{f_1f_2}_{ab;i} \otimes \Gamma^{f_1f_2}_{cd;i}\right)
\end{equation}
where the $H_i$ parametrise the strengths and the profiles of the non-local
interactions and $f_i$ labels the flavor (either $q$ or $Q$), $\tilde \Gamma = \gamma_0 \Gamma^\dagger \gamma_0$. The Dirac and flavor structures for the present model are then
\begin{eqnarray}
\Gamma^{qq}_{\{S,P,V,A\}} &=& \left\{ 1, i\gamma_5\tau^a, \gamma_\mu\tau^a, 
\gamma_\mu\gamma_5\tau^a \right \}, \nonumber\\
\Gamma^{qQ}_{\{S,P,V,A\}} &=& \left\{ 1, i\gamma_5, \gamma_\mu, \gamma_\mu\gamma_5\right \}, \nonumber\\
\Gamma^{QQ}_V &=& \gamma_\mu. 
\end{eqnarray}
Chiral symmetry then imposes that $H^{\{qq,qQ\}}_S = H^{\{qq,qQ\}}_P$ and
 $H^{\{qq,qQ\}}_V = H^{\{qq,qQ\}}_A$. This choice  
can be explicitly checked by using the global transformation for the 
light quark field \cite{bou08} and remembering that the
heavy quark field is invariant under this symmetry.

The most
general form of $H$s is constrained by translational invariance. To make this property explicit
a change of variables as in Ref.~\cite{Nog07} is made, namely
\begin{eqnarray}
X = \frac{1}{2} \left(-x_1-x_2+x_3+x_4\right)&,& \qquad X' = \frac{1}{4}
\left(x_1+x_2+x_3+x_4\right), \nonumber \\
x = x_2-x_1&,& \qquad x' = x_4-x_3.
\end{eqnarray}
With these and suppressing all indices, the kernels become
\begin{eqnarray}
H^{f_1f_2}_i(x_1,x_2,x_3,x_4) &=& \prod_k \int dp_k e^{-i\sum_j x_j \cdot p_j}
H^{f_1f_2}_i(p_1,p_2,p_3,p_4) \nonumber \\
&=&  \prod_k \int dp_k e^{i\left(p_1-p_2\right)\cdot
\frac{x}{2}}e^{i\left(p_3-p_4\right)\cdot \frac{x'}{2}} 
e^{i\left(p_1+p_2-p_3-p_4\right)\cdot \frac{X}{2}} 
\nonumber \\
&\times& e^{-i\left(p_1+p_2+p_3+p_4\right)\cdot X'} H^{f_1f_2}_i(p_1,p_2,p_3,p_4)
\end{eqnarray}
where the momenta are taken to be in-going. Translation invariance then amounts 
to requiring that under an arbitrary shift by a four vector $a$, the kernels respect
\begin{eqnarray}
H^{f_1f_2}_i(x_1+a,x_2+a,x_3+a,x_4+a) =  H^{f_1f_2}_i(x_1,x_2,x_3,x_4),
\end{eqnarray}
or more specifically that the $H_i$ do not depend on $X'$. This then restricts
their form in momentum-space to 
\begin{equation}
\tilde H^{f_1f_2}_i(p_1,p_2,p_3,p_4) =\left(2\pi\right)^4 \delta^{(4)}(P')
\tilde H^{f_1f_2}_i(p',p,P)
\end{equation}
where we have defined 
\begin{eqnarray}
\qquad P = \frac{1}{2}\left(p_1+p_2-p_3-p_4\right) &,\qquad& P' = \left(p_1+p_2+p_3+p_4\right)
\nonumber \\
p = \frac{1}{2}\left(p_1-p_2\right)&,\qquad&   p' = \frac{1}{2}\left(p_3-p_4\right).
\label{momenta}
\end{eqnarray}
The original NJL model is given by 
\begin{equation}
\tilde H^{f_1f_2}_i(p_1,p_2,p_3,p_4) = G^{f_1f_2}_i\left(2\pi\right)^4 \delta^{(4)}(P')
\end{equation}
where the $G^{f_1f_2}_i$ are the interaction strengths, which have a dimension of inversed
energy squared. 
Here, we will consider a fully-separable interaction in momentum-space 
inspired by an instanton-based approach \cite{Pla98}, i.e.,
\begin{eqnarray}
\tilde H^{f_1f_2}_i(p_1,p_2,p_3,p_4) = \frac{1}{2}(2\pi)^4 G^{f_1f_2}_i \delta(P') f_{f_1}(p_1)
f_{f_1}(p_2)f_{f_2}(p_3)f_{f_2}(p_4)
\label{K}
\end{eqnarray}
where $f(p_i)$ are the quark form factors modeling the non-locality of the interactions. 
These are normalised to one at zero impulse and their specific forms will be
chosen, in the light sector, to provide confinement. Moreover, they act as UV
regulators for the loop integrals removing the need for a UV cutoff as in the original NJL
model.
The $G_i$ constants scale like $1/N_c$, which can be inferred by considering the
simplest four--point interaction in QCD : the one-gluon exchange interaction between four
quarks. The $1/N_c$ scaling of the four--point vertex will be used in what
follows to determine an approximation scheme consistent with chiral symmetry. 
This particular choice of kernels also greatly simplifies the search for 
mesonic bound states and facilitates numerical integration. 

\section{Quark propagators}

\subsection{Light quark sector}
\label{light_quark_sector}
 The general solution of the Schwinger-Dyson equation (SDE) in
 momentum-space is given by \cite{Rob94}:
\begin{equation}
S_q(p) = Z_q(p) \frac{\displaystyle{\not} p -m_q(p)}{p^2 -m_q^2(p)}
\end{equation}
where $m_q$ and $Z_q$ are the momentum-dependent mass function and
wavefunction renormalization respectively.


Working in the mean field approximation or equivalently at leading order in $1/N_c$, 
the light quark propagator  for the action of
Eq.~(\ref{LENJL_interactions}) reduces to \cite{Pla98,Dmi95}
\begin{equation}
 S_q(p) = \frac{1}{\displaystyle{\not} p -m_q(p)}
 \label{propagator}
 \end{equation}
 where the dynamical mass is given by 
 \begin{equation}
 m_q(p) = m^q_c + iG_S f^2_q(p)\int \frac{d^4k}{(2\pi)^4}  f^2_q(k){\rm
 Tr} \left\{S_q(k) \right\}.
 \label{mq3}
 \end{equation}
 We note that only the scalar channel gives a non-zero contribution and the 
 dynamical mass scales like $N_C^0$ as in QCD \cite{Dmi95}.
 The above gap equation then
 admits the solution \cite{Pla98}
 \begin{equation}
 m_q(p) = m^q_c + (m_q(0)-m^q_c)f_q^2(p)
 \label{mq}
 \end{equation}
 where $m_q(0)$ is the dynamical mass at zero momentum. The quark propagator can also be directly
 linked to the quark condensate in the chiral limit through the expression
  \begin{eqnarray}
 \left<\bar q q\right>_0 &=& -i \int \frac{d^4k}{(2\pi)^4} Tr\left[S_q(k)\right]. 
 \label{quark_condensate}
 \end{eqnarray}

From the quark propagator it can be readily inferred that there will be no poles
on the real axis in two cases: either the poles are complex \cite{Bla01} or there are no
poles at all in the complex plane \cite{Rad04}. Here, we will consider the second case. 
However, to illustrate how this property is manifest we need to analytically continue 
the quark propagator to Euclidean space, i.e., the denominator becomes $p_E^2 + m^2(p)$.


We then follow Ref.~\cite{Rad04} where the inverse of the quark propagator
 denominator, in the chiral limit, is parametrized as
\begin{equation}
\frac{1}{p^2 + m_q^2(p)} = \frac{1-e^{-\mu p^2}}{p^2}.
\end{equation}
Alternatives also exist such as the one found in Ref.~\cite{Bub92}.
Various limits can then be considered. For large positive $p^2$, this quantity
behaves as expected perturbatively, i.e., $1/p^2$.
This is not the case for large {\it negative} $p^2$ where it diverges. This
is not a major problem provided the mass function is probed only for small negative 
$p^2$, i.e., small time--like separation in Minkowski space. 
As the infrared limit is approach, i.e., $p^2 \rightarrow 0$,  the inverse quark
propagator becomes constant and equal to $\mu$. The cutoff parameter is then 
seen to be equal to $\mu = 1/m^2(0)$. This is sufficient to show that no poles
exist. Mathematically, this function property is called entire, and  
this realization of quark confinement is deemed analytic.
Reinstating the current quark mass yields
\begin{equation}
m_q(p) = \sqrt{\frac{m^{q2}_{c} +
p^2e^{-\mu\left(p^2+m^{q2}_{c}\right)}}{1-e^{-\mu\left(p^2+m^{q2}_{c}\right)}}}
\label{mq_2}
\end{equation}
where the principal branch has been chosen.\footnote{This choice, through Eq.(\ref{mq}), leads to an action
that is not linear in the current
mass, contrary to the QCD action, and results from enforcing analytic confinement in this particular model. However, it can be shown that the GMOR relation holds to first order in
$m_c$ for such a mass model (see Ref.\cite{Rad04}).}  In what follows, we will refer to the model
using this functional form for the dynamical mass as Model I (MI). This choice will complicate the 
evaluation of $n$--point functions as branch cuts will generally appear. 
Since the zero--momentum light--quark mass should be of the order of a few
hundred MeVs \cite{Won00,Iva05}, the parameter $\mu$ will be greater than one.
Therefore, for $p^2 > 0$, the dynamical mass will exhibit a steep
decrease. To control this ultraviolet behavior, we also consider an  alternative dynamical mass model (MII), which is constructed
by substituing $m_c^q \rightarrow m_c^q +
(m_0-m_c^q)e^{-\alpha p^2}$ in Eq.~(\ref{mq_2}) where $m_0$ and $\alpha$ are two
additional parameters. The steep decline for $p^2>0$ can then
be overcome.

\subsection{Heavy quark sector}
For the heavy flavor sector, again no wavefunction renormalization is possible for this
model at leading order in $1/N_c$. Furthermore, there is no heavy quark scalar interaction in the quark
action. Therefore, there is no dynamical mass
generation at the mean field level contrary to the light quark sector.
This leaves the charm quark form factor unspecified. Here, we will chose: 
\begin{equation}
f_Q(p) = e^{-\beta_Q p^2}
\end{equation}
in Euclidean space where $\beta_Q$ will be fixed by considering the decay of the $J/\psi$ into dileptons
(see Appendix \ref{Decays}).

\section{Meson bound states}

\subsection{Meson-quark vertex functions and meson propagators}
Interactions amongst constituents lead to the emergence of bound states
provided attractive channels exist between them.
In QCD, these occur, at least perturbatively, in the singlet-color channel
between a quark and anti-quark. At all orders in
quantum field theory, bound states are found by considering the pole structure of the
scattering matrix, $S$.
Following Refs.~\cite{Bet51,Alk01}, the inhomogeneous Bethe-Salpeter equation (BSE) satisfied by
the $T$--matrix for a given flavor content, i.e.,  the interacting part of the scattering matrix,  is
\begin{equation}
T(p,p',P) = K(p,p',P) + \int dp^{\prime\prime} K(p,p^{\prime\prime},P)S(p_-)T(p^{\prime\prime},p',P)S(p_+)
\label{BSE2}
\end{equation}
where the Dirac indices and flavor labels have been suppressed for clarity and $p_\pm = p^{\prime\prime}\pm\frac{1}{2}P$.

To solve the BSE, we have to specify, besides the quark propagators, the scattering kernels. In general, 
since they comprise of an infinite sum of diagrams, a truncation scheme must be implemented. 
If the Hartree propagator for light quark is used 
and we require chiral symmetry to be maintained, then the scattering kernels involving light quarks
are uniquely determined: they cannot contain exchange terms 
and their functional forms are fixed to that of Eq.~(\ref{K}).
This is known in low-energy nuclear physics as the random-phase approximation (RPA) and is the
counterpart of the mean field hypothesis for solving the SDE. Moreover, without any surprises,
their $N_C$--scaling is that of the $G_i$ couplings. The kernels being fully 
separable, the BSE admits the solution
\begin{equation}
T^{f_1f_2}\left(p,p',P\right)  = f_{f_1}(p_1)f_{f_1}(p_2)f_{f_2}(p_3)f_{f_2}(p_4)
\hat T^{f_1f_2}(P)
\end{equation} 
where momentum conservation is implicit and $\hat T^{f_1f_2}(P)$ are the $T$--matrix with the $f(p)$
factored out. 

Next following Ref.~\cite{Kli90}, the BSE is decomposed into independent
Dirac channels. This can be achieved by using the projection operators
\begin{equation}
T_{\mu\nu} = g_{\mu\nu} - \hat P_\mu \hat P_\nu, \qquad L_{\mu\nu} =
\hat P_\mu \hat P_\nu\qquad
\end{equation}
where $\hat P_\mu = P_\mu/\sqrt{P^2}$. The scattering kernels can
then  be re-arranged for the pseudo-scalar and scalar channels as 
\begin{equation}
\hat K^{f_1f_2}_{\{S,P\}} = G^{f_1f_2}_{\{S,P\}}\left(\tilde \Gamma^{f_1f_2}_{\{S,P\}} \otimes
\Gamma^{f_1f_2}_{\{S,P\}} \right) + 
G^{f_1f_2}_{\{V,A\}} \left(\tilde \Gamma^{f_1f_2;L}_{\{V,A\}} \otimes \Gamma^{f_1f_2;L}_{\{V,A\}}\right)
\end{equation}
and 
\begin{eqnarray}
\hat K^{f_1f_2}_{\{V,A\}} = G^{f_1f_2}_{\{V,A\}}\left(\tilde \Gamma^{f_1f_2;T}_{\{V,A\}} \otimes
\Gamma^{f_1f_2;T}_{\{V,A\}} \right)
\end{eqnarray}
for the vector and axial ones
where 
\begin{equation}
\Gamma^\mu_{\{V,A\}} = \left(T^\mu_\nu+L^\mu_\nu\right)\Gamma_{\{V,A\}}^\nu =
\Gamma^{T\mu}_{\{V,A\}} + \Gamma^{L\mu}_{\{V,A\}}.
\label{tensor}
\end{equation}
Note also that the left-- and right--hand parts of a given flavor and Dirac matrix
product is associated with the in-- and out--going quark states, respectively. Thus
the kernels should be read from left to right. 
Similarly, the $\hat T$--matrix can be decomposed into products of Lorentz
covariant tensors giving
\begin{eqnarray}
\hat T^{f_1f_2}_{\{S,P\}} &=& M^{f_1f_2}_{\{SS,PP\}}\left(\tilde \Gamma^{f_1f_2}_{\{S,P\}}
\otimes  \Gamma^{f_1f_2}_{\{S,P\}} \right) + 
M^{f_1f_2}_{\{SV,PA\}} \left(\tilde \Gamma^{f_1f_2}_{\{S,P\}} \otimes
\Gamma^{f_1f_2;L}_{\{V,A\}} \right)  \nonumber \\
&+& M^{f_1f_2}_{\{VS,AP\}} \left( \tilde \Gamma^{f_1f_2;L}_{\{V,A\}}  \otimes 
\Gamma^{f_1f_2}_{\{S,P\}} \right) +
M^{f_1f_2;L}_{\{VV,AA\}} \left(\tilde  \Gamma^{f_1f_2;L}_{\{V,A\}} \otimes 
\Gamma^{f_1f_2;L}_{\{V,A\}}\right), 
\end{eqnarray}
and 
\begin{eqnarray}
\hat T^{f_1f_2}_{\{V,A\}} &=& M^{f_1f_2;T}_{\{VV,AA\}} \left(\tilde \Gamma^{f_1f_2;T}_{\{V,A\}}\otimes 
\Gamma^{f_1f_2;T}_{\{V,A\}}\right)
\end{eqnarray}
where the $M$ components are functions of $P^2$. It is clear that, in general, 
mixing occurs between channels, i.e., the longitudinal component of the
vector and axial channels contribute to the scalar and pseudo-scalar ones
respectively, and that the $T$--matrix is then block--diagonal. 
From the BSE, it is then seen
that for a given flavor and Dirac channels, a given $M$ matrix satisfies the equation
\begin{equation}
M = G\left[1+
JM\right] = \frac{G}{1-GJ} 
\label{sol_M}
\end{equation}
where all labels have been suppressed. The fermion--loop matrix is then given by
\begin{equation}
iJ^{f_1f_2}_{ij}(P^2) = -\int dk f^2_{f_1}(k_-) f^2_{f_2}(k_+) {\rm Tr}\left[
\Gamma^{f_1f_2}_{i} S_{f_1}(k_-)
\tilde\Gamma^{f_1f_2}_{j} S_{f_2}(k_+)\right]
\label{fermion_loop}
\end{equation}
where the minus sign on LHS is due the normal ordering of fermion fields, $k_\pm = k \pm \frac{P}{2}$, and
the matrix indices are $i$ and $j$.\footnote{The analytic continuation for these two-loop integrals is
explained in Appendix \ref{analytic}.}
This equation shows that a pole will develop in
the $T$--matrix
when 
\begin{equation}
\Delta = \left|1-GJ(m_M^2)\right| = 0
\label{pole}
\end{equation}
where $m_M$ is the mass of the meson. On one hand, near this point, the $T$--matrix admits the solution
\begin{equation}
T(p,p',P) \approx  \frac{ i\bar \chi_M(p,P) \otimes i\chi_M(p',P) }{P^2-m_M^2} 
\label{T_near_pole}
\end{equation}
where the wavefunction for meson $M$ is defined as
\begin{equation}
\chi_M(p',P) = \left<\bar q_{f_1} q_{f_2}|M(P)\right> =  g_M f_{f_1}(p_1)f_{f_2}(p_2)
\left(1-\frac{a_M}{m_M}\displaystyle{\not}\hat P\right)\Gamma^{f_1f_2}_M
\end{equation}
with $p^\prime = \frac{1}{2}\left(p_3-p_4\right)$.
The coupling and mixing parameters are found by solving Eq.~(\ref{sol_M}) using 
Eq.(\ref{T_near_pole}) yielding for a meson of spin $s$: $g_m = (-1)^s M^{00}/d\Delta/dP^2$ and $a_m = M^{11}/M^{00}$.
Note that for channels where
there is no mixing $a_M =0$. On the other hand, for $P^2$ far from the on-shell condition, the $\hat T$--matrix can be written as
\begin{equation}
\hat T_M = ig_M\hat \Gamma^i_M \otimes ig_M\hat \Gamma^j_M \mathcal{D}_M(P^2)
\label{T_off_shell}
\end{equation}
where $\hat \Gamma^i_M = \left(1,\displaystyle{\not}\hat P\right)\otimes \Gamma^{f_1f_2}_M$ and with the meson propagator given by
\begin{equation}
\mathcal{D}_{M}(P) = -\frac{1}{g_M^2} M_M(P)
\end{equation}
where the tensorial structure is suppressed.

\subsection{Meson interactions}
Having discussed the couplings between quarks and mesons, the
interactions amongst the mesons are now examined. As for the meson self--energies,
only interactions mediated by quarks will be studied; mesonic fluctuations being
sub-leading in a $1/N_c$ expansion.
\begin{figure}[!htb]
\begin{center}
\begin{tabular}{cc}
\scalebox{1.00}[1.00]{\includegraphics{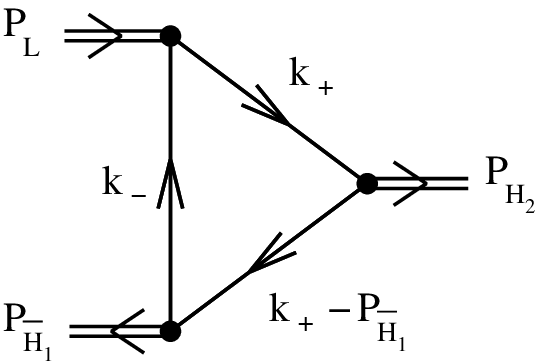}}&
\scalebox{1.00}[1.00]{\includegraphics{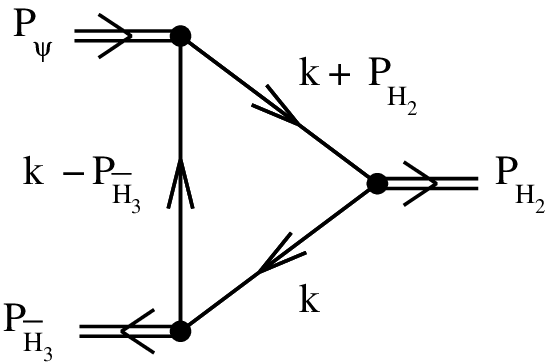}}
\end{tabular}
\caption{Momentum conventions for the three--point vertex functions.}
\label{triangle_diagrams}
\end{center}
\end{figure}

To evaluate the dissociation cross sections, the three-- and four--point
interactions have to be written down. The former are further
divided into two, namely interactions between one light meson, either the pion or the
$\rho$ meson and two open charmed mesons; or interactions between the $J/\psi$ and two
open charmed mesons. The momentum flows for these two cases are depicted in
Fig.~\ref{triangle_diagrams}. Here, in general, one of the three mesons will be off-shell and the
kinematical variable $t$ is then the associated momentum transfer. The expressions for the meson form factors
are
\begin{eqnarray}
iF^i_{L\bar H_1(H_2)}(t) &=& -\int dk {\rm Tr}\left[i\chi_LiS_q\left(k_-\right)
ig_{H_2}\hat\Gamma^i_{H_2}iS_Q
\left(k_+-P_{\bar H_1}\right)i\bar \chi_{\bar H_1}
iS_q\left(k_+\right)\right], \nonumber \\ \\
iF^i_{LH_2(\bar H_1)}(t) &=& -\int dk {\rm Tr}\left[i\chi_LiS_q\left(k_-\right)i\bar
\chi_{H_2}iS_Q
\left(k_+-P_{\bar H_1}\right)ig_{H_1}\hat\Gamma_{\bar H_1}
iS_q\left(k_+\right)\right]\nonumber \\
\end{eqnarray}
and
\begin{eqnarray}
iF^i_{\psi H_2(\bar H_3)}(t) &=& -\int dk {\rm Tr}\left[i\chi_\psi
iS_Q\left(k-P_{\bar H_3}\right)
ig_{H_3}\hat\Gamma^i_{H_3}iS_q(k)i\bar
\chi_{H_2}iS_Q\left(k+P_{H_2}\right)\right] \nonumber \\ \\
iF^i_{\psi \bar H_3(H_2)}(t) &=& -\int dk {\rm Tr}\left[i\chi_\psi
iS_Q\left(k-P_{\bar H_3}\right)
i\bar \chi_{H_3}iS_q(k)i
g_{H_2}\hat\Gamma^i_{H_2}iS_Q\left(k+P_{H_2}\right)\right] \nonumber \\
\label{JH1H2}
\end{eqnarray}
where the momentum arguments of the wavefunctions have been suppressed for clarity, 
$L$ and $H$ label the light and open charmed mesons respectively, the minus sign
is due to the fermion-loop, and the round parentheses indicate which meson is off-shell. The extra
label $i$ notes that the form factor can be a two-component vector due to mixing.

\begin{figure}[!htb]
\begin{center}
\scalebox{1.00}[1.00]{\includegraphics{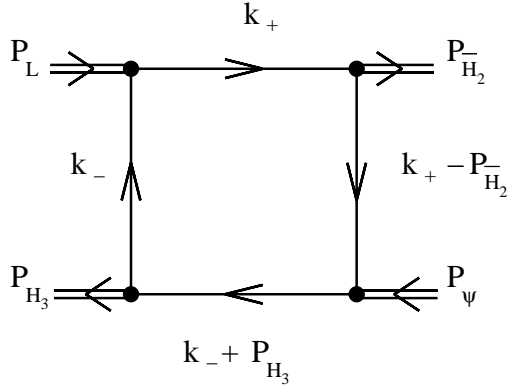}}
\caption{Momentum convention for the four--point vertex functions.}
\label{box_diagram}
\end{center}
\end{figure}
Similarly, the general expression for the four--point interaction, illustrated in
Fig.~\ref{box_diagram}, is given by
\begin{eqnarray}
iF_{L\psi\bar H_2H_3}(s,t) = -\left.\int dk
{\rm Tr}\right[&i&\chi_LiS_q\left(k_-\right)i\bar
\chi_{ H_3}iS_Q
\left(k_-+P_{ H_3}\right)i\chi_{\psi}\nonumber \\
&i&S_Q\left(k_+-P_{\bar
H_2}\right)\left.i\bar \chi_{\bar H_2}iS_q\left(k_+\right) \right] 
\end{eqnarray}
where again the minus sign is due to the fermion-loop. The analytic continuation of the loop integrals is
discussed in Appendix~\ref{analytic}. 


\section{Results}

\subsection{Parameter fixing}
The isospin-averaged masses of the pion, $\rho$, $D$, $D^*$, and $J/\psi$ mesons
used in what follows are $0.138\,\mbox{GeV}$, $0.770\,\mbox{GeV}$, $1.868\,\mbox{GeV}$, $2.009\,\mbox{GeV}$, 
and $3.096\,\mbox{GeV}$ respectively. The first model (referred as MI) is that of Eq.~(\ref{mq_2}) and has four parameters, namely 
$m_c^Q$, $\beta_Q$, $\mu$, and $m_c^q$. Three observables, i.e, the pion decay constant
[Eq.~(\ref{f_pi_2})], the dielectron decay width of the $J/\psi$ meson
[Eq.~(\ref{gamma_J_epem})], and the the decay width of  the $D^*$ into  $D$--$\pi$
final state [Eq.~(\ref{g_piDsD})], can be then used to partly constrain the possible parameter values.
Specifically, $m_c^q$ is determined in order to reproduce the pion decay constant
value since its expression [Eq.~(\ref{f_pi_2})] 
has a strong $m_c^q$--dependence. While, for a fixed $m_c^Q$, 
the range parameter, $\beta_Q$, is constrained by the experimental 
dielectron decay width of the $J/\psi$ [Eq.(\ref{gamma_J_epem})]. This leaves the open charmed decay
constant [Eq.~(\ref{g_piDsD})] to fix both $m^Q_c$ and $\mu$. To restrict the possible
solutions, we also required that the light-quark dynamical mass at zero-momentum
be of the order of a few hundred MeVs as it is seen from lattice results \cite{Nog07} and as used
in other phenomenological studies \cite{Won00,Iva05}. We further require that $m_c^Q$ obey the
constraint $m_{J/\psi} < 2m_c^Q$ since the model is
non-confining in the heavy sector. With these additional constraints, one possible
set of values is $m_c^Q = 1.59\,\mbox{GeV}$, $\beta_Q = 0.06\,\mbox{GeV}^{-2}$, $m_c^q = 12\,\mbox{MeV}$, and  $\mu
= 8 \,\mbox{GeV}^{-2}$. These then yield $f_\pi = 94.78\,\mbox{MeV}$, $\Gamma_{J/\psi \rightarrow e^+ +e^-} =
5.44\,\mbox{keV}$, and $g_{\pi D D^*} = 18.62$
which are to be compared to the experimental values of $f_\pi = 93\,\mbox{MeV}$, $\Gamma_{J/\psi \rightarrow e^+ +
e^-} =  5.5 \pm 0.14 \pm 0.02\,\mbox{keV}$ \cite{Yao06}, and $g^{\mbox{exp}}_{\pi DD^*} = 17.9\pm 0.3\pm 1.9$  
\cite{Ana03}. The resulting light--quark mass at zero momentum is then $m_q(0) = 0.354\,\mbox{GeV}$, while the quark
condensate is $-(237)^3\,\mbox{MeV}^{3}$ as calculated using Eq.~(\ref{quark_condensate}). Both these
values are slightly higher than those found in Ref.~\cite{Pla98}.

Introducing the substitution discussed earlier in Section \ref{light_quark_sector} into the
Eq.~(\ref{mq_2}), the number of parameters in the
light sector increases by two (model MII). Additional information is thus
needed. An alternative could be to calculate other observables such as the $\rho$ decay into two
pions in order to fix the extra parameters. Rather, here a parametrisation of lattice data is used.\footnote{The
quark mass is not a gauge-invariant quantity and thus dependent on the gauge
chosen to carry out the simulation. The purpose here is to capture some flavor of the
QCD dynamics.}
Specifically, we will utilize the one proposed in Ref.~\cite{Nog07}, namely
\begin{equation}
m_{q}\left(p\right) = \alpha_m
\left(\frac{\Lambda_m^2}{\Lambda_m^2+p^2}\right)^{\frac{3}{2}}
\end{equation}
where $\alpha_m = 0.343 \,\mbox{GeV}$ and $ \Lambda_m = 0.767\,\mbox{GeV}$.
 Fitting our model [Eq.~(\ref{mq_2})] to this parametrisation for $p^2 \in \left[0,1\right]$ gives 
 $m_0 = 0.227\,\mbox{GeV}$, $\alpha = 1.096 \,\mbox{GeV}^{-2}$ and $\mu =
11.786\,\mbox{GeV}^{-2}$. The light current mass is then taken to be $m^q_c =
5.5\,\mbox{MeV}$ yielding $f_\pi = 92.11\,\mbox{MeV}$. The zero-momentum
dynamical mass is then $0.331\,\mbox{GeV}$, while the quark condensate is now
$-(239)^3\,\mbox{MeV}^{3}$. With the same values for the heavy--sector parameters, 
the coupling constant $g_{\pi DD^*}$ is $18.35$ which is a prediction of 
the model and well within the experimental tolerance \cite{Ana03}. 
\begin{figure}[htb]
\begin{center}
\scalebox{0.45}[0.45]{\includegraphics{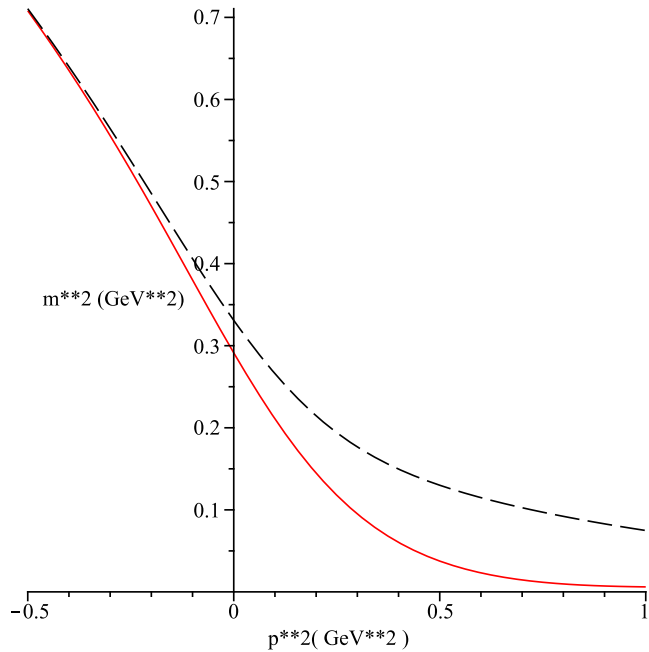}}
\caption[(Color online) Light--quark dynamical mass models.]
{(Color online) Light--quark dynamical mass models with $u=p^2$. The solid and dashed lines are the MI and MII
models respectively.}
\label{comp_mass_model}
\end{center}
\end{figure}

Fig.~\ref{comp_mass_model} shows a comparison between the two light-quark mass
models discussed. As seen, for $p^2 \leq 0 $, the two models are very similar. The
main difference is for $p^2 > 0$ where  for MI the dynamical mass drops very quickly
down to its asymptotic value of $m_c^q$. Both models have similar zero-momentum mass,
quark condensate, and calculated observables. The only noted difference is in
the light-quark current mass for which there is more than a factor two
difference between models. Although the current mass for MII is within the range
given by PDG \cite{Yao06}, i.e., $1.3$ to $5$ MeV for the $u$--quark and of $3$ to $7$ MeV for the
$d$--quark, one could wonder if introducing two extra parameters, $m_0$ and
$\alpha$,  in order to reduce the current quark mass value is justified. At this point, 
we will retain the two models to ascertain if any other differences occur for
vertices and cross sections. 

With the parameters for MI and MII, the four-quark couplings, the meson-quark couplings, and
the mixing coefficients can be evaluated. Their values are listed in Table~\ref{tab_couplings}.  
Since the $D^*_0$ and $D_1$ are the chiral partners of the $D$ and $D^*$ respectively,
the masses are not independent and have to be calculated by finding the zeros of the respective
meson propagator denominators. Doing so yields
$m_{D^*_0} = 2.064\,\mbox{GeV}$ and $m_{D_1} = 2.249\,\mbox{GeV}$,
and, $m_{D^*_0} = 2.045\,\mbox{GeV}$ and $m_{D_1} = 2.231\,\mbox{GeV}$ for MI
and MII, respectively. These are to be
compared to the experimental masses of $m_{D^*_0} = 2.40$
GeV and $m_{D_1} = 2.43$ GeV.
It is quite clear that neither model is capable of reproducing the
absolute masses and the mass difference, i.e., $\Delta m_{exp} =
0.03\,\mbox{GeV}$. This
problem then implies that the interaction kernels or, in non-relativistic terms the
potentials, require further modeling. This is left for a future study.  
\begin{table}[h]
\begin{center}
\begin{tabular}{|c|c|c||c|c|c|}
\hline
$G_{M}$ & MI & MII  & $g_M$ & MI & MII \\
\hline
$G_{J/\psi}$ & -1.145 & -1.145 & $g_{J/\psi}$ & 1.717 & 1.717 \\
\hline
$G_{D^*}$ & -6.690 & -5.257 & $g_{D^*}$ & 2.025 & 1.842\\
\hline
$G_{D_1}$ & -6.690 & -5.257 & $g_{D_1}$ & 1.955 & 1.772\\
\hline
$G_D$ & 17.661 & 11.977 & $g_{D}(a_D)$ & 4.667(0.301) & 4.166(0.315)\\
\hline 
$G_{D_0^*}$ & 17.661 & 11.977 & $g_{D^*_0}(a_{D^*_0})$ & 3.654(0.166) & 3.828(0.205) \\
\hline
$G_\rho$ & -7.070 & -6.147 & $g_{\rho}$ & 1.336 & 1.219\\
\hline
$G_\pi$ & 52.562 & 31.052 & $g_{\pi}(a_\pi)$ & 3.768(0.0220) & 3.615(0.0233)\\
\hline
\end{tabular}
\end{center}
\caption{Quark-quark couplings, meson-quark couplings, and mixing coefficients.}
\label{tab_couplings}
\end{table}

\subsection{Meson propagators and vertices}
Having fixed the parameters for the two models, the meson propagators and vertices can now be examined. 
For the meson propagators, $\mathcal{D}_{M}$, as can be seen from Eq.~(\ref{sol_M}), the asymptotic behavior is controlled
by the the two-point functions, $J_M$, as the quark-quark coupling, $G_M$, is independent of momentum.\footnote{ For simplicity, mixing is ignored here. Adding it does not
alter the conclusions.} As $t\rightarrow \infty$, $J_M \rightarrow \infty$ and
$\mathcal{D}_{M} \rightarrow 0$, while for $t\rightarrow -\infty$, $J_M \rightarrow 0$ and
$\mathcal{D}_{M} \rightarrow -\frac{G_M}{g_M^2}$. Near the meson pole, the scalar part of the corresponding propagator
[Eq.~(\ref{T_near_pole})] is expected to behave as 
\begin{equation}
\mathcal{D}^{\mbox{pole}}_{M} (t) \propto \frac{(-1)^{s}}{t-m_M^2}
\label{lagrangian_prop}
 \end{equation}
 where $s$ is the meson spin. Note that this form is used for the
 phenomenological Lagrangian studies of
 Refs~\cite{Mat98,Lin00,Lin00_2,Hag00,Hag01,Oh01,bou08}.
\begin{figure}[!htb]
\begin{tabular}{cc}
\hspace*{0cm}
\scalebox{0.326}[0.326]{\includegraphics{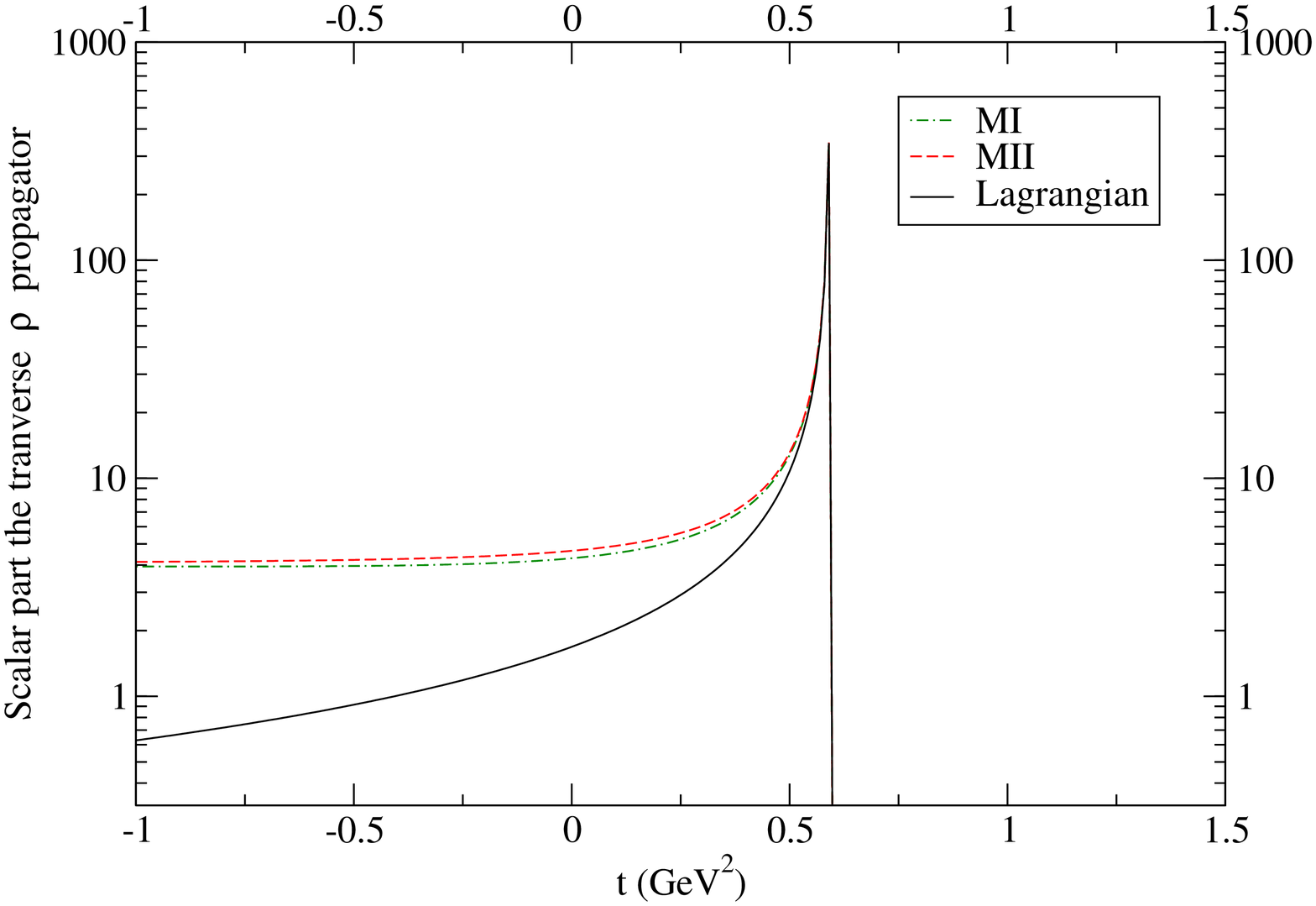}} &
\hspace*{-0.5cm}
\scalebox{0.326}[0.326]{\includegraphics{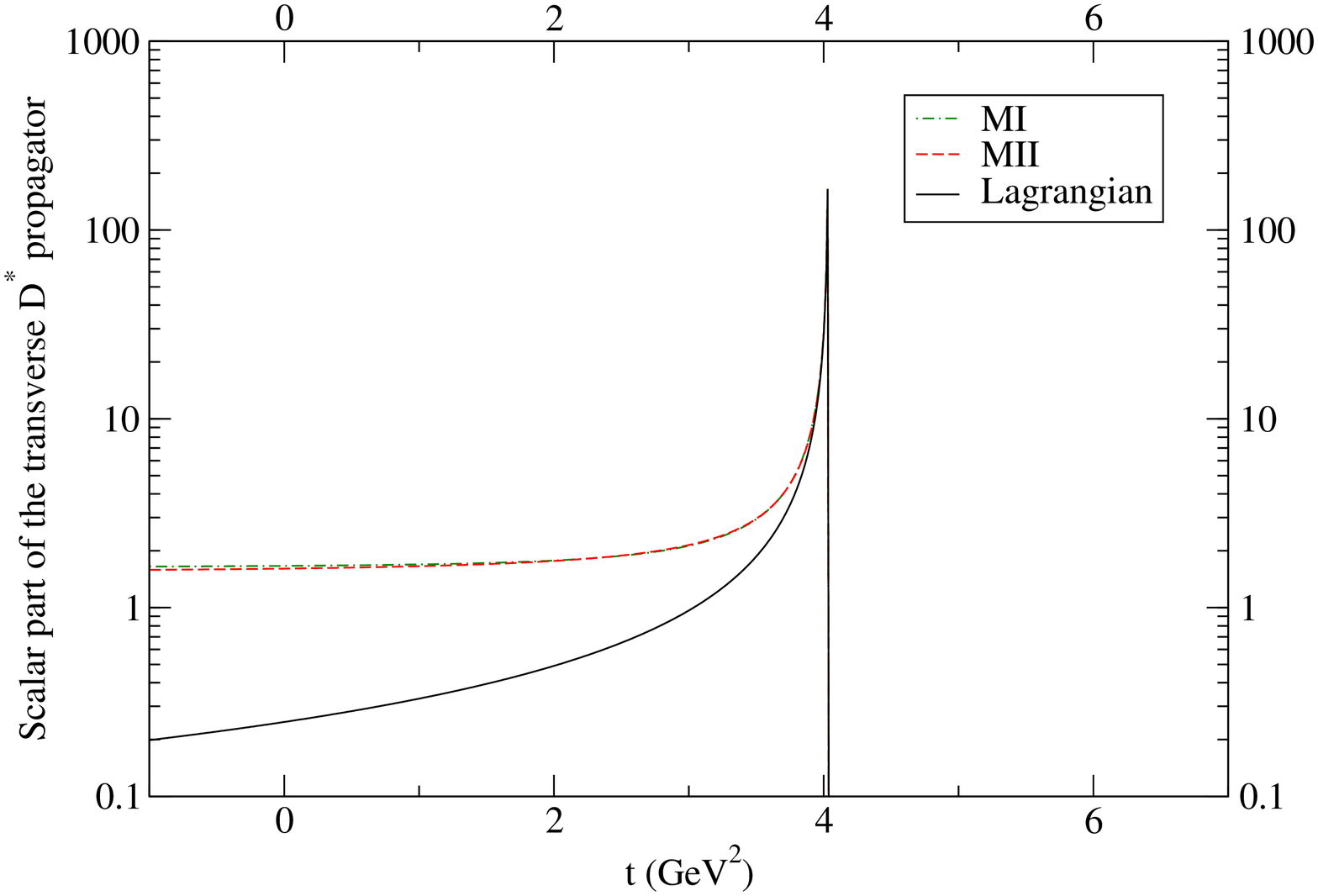}} 
\end{tabular}
\caption[(Color online) Scalar parts of the transverse $\rho$ and $D^*$ meson propagators.]
{(Color online) Scalar parts of the transverse $\rho$ and $D^*$ meson propagators near their respective pole.}
\label{meson_propagators}
\end{figure}

 Fig.~\ref{meson_propagators} \footnote{A logarithmic scale was used in order to permit the discrimination
 between the three curves. At the pole, the divergence should be infinite and the appearance of
 the contrary is just an artifact of the finite number of points in the numerical evaluation of the
 propagator. } illustrates the behaviors of the $D^*$ and $\rho$ propagators for MI and MII
 compared to that of Eq.~(\ref{lagrangian_prop}).  It is quite clear that the
 phenomenological propagators are comparable to the NJL ones only near the poles, and that the
 differences between the propagators of MI and MII are not significant. Although not shown on the figure, this 
 latter observation holds true for $t > m_M^2$. Note that for $t\rightarrow -\infty$ the
 NJL-propagators exhibit the expected asymptotic behavior, i.e., it is non--zero as for the Lagrangian one,
 but rather $-\frac{G_M}{g_M^2}$.

\begin{figure}[!htb]
\begin{tabular}{cc}
\hspace*{0cm}
\scalebox{0.326}[0.326]{\includegraphics{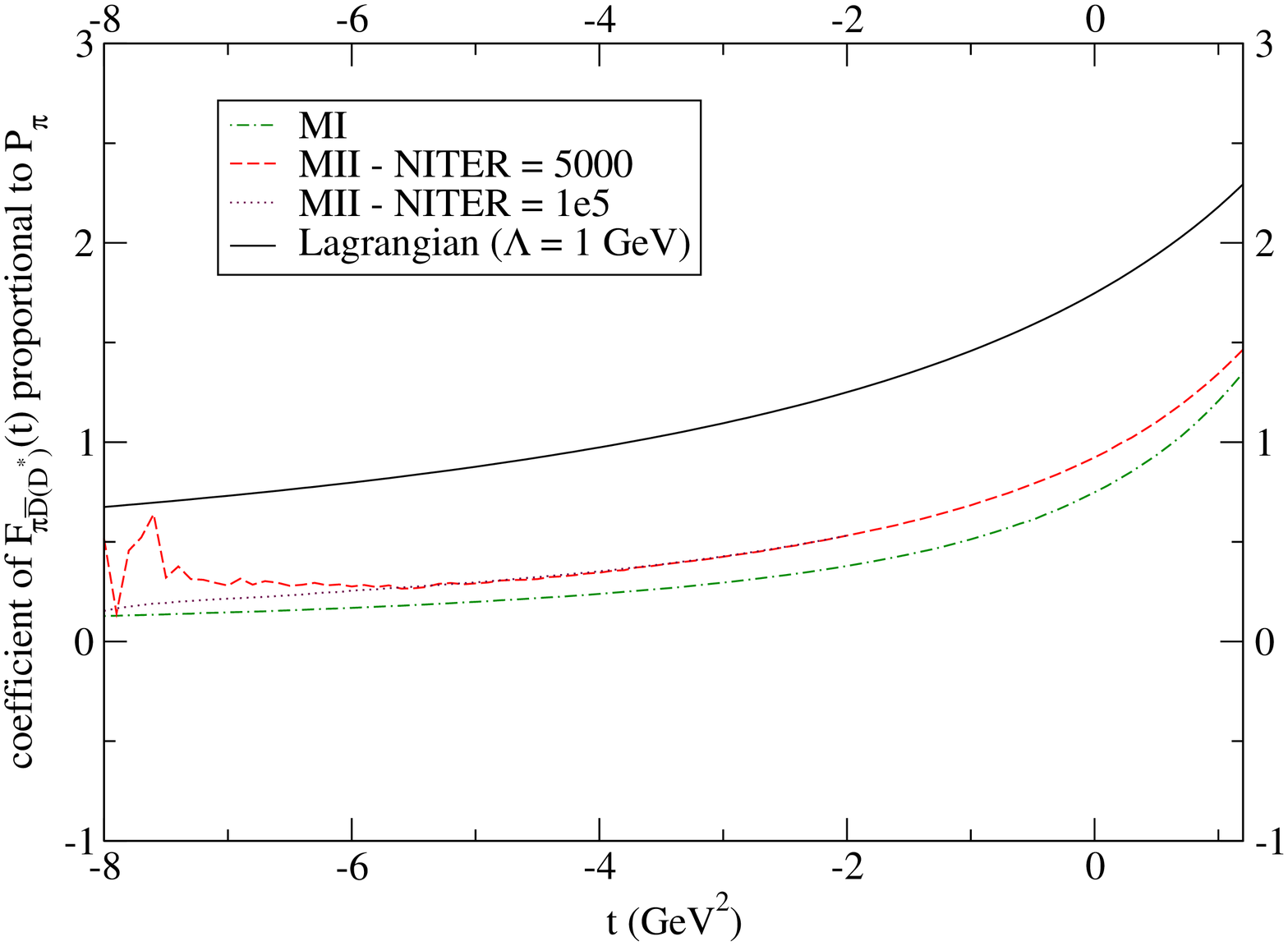}} &
\hspace*{-0.5cm}
\scalebox{0.326}[0.326]{\includegraphics{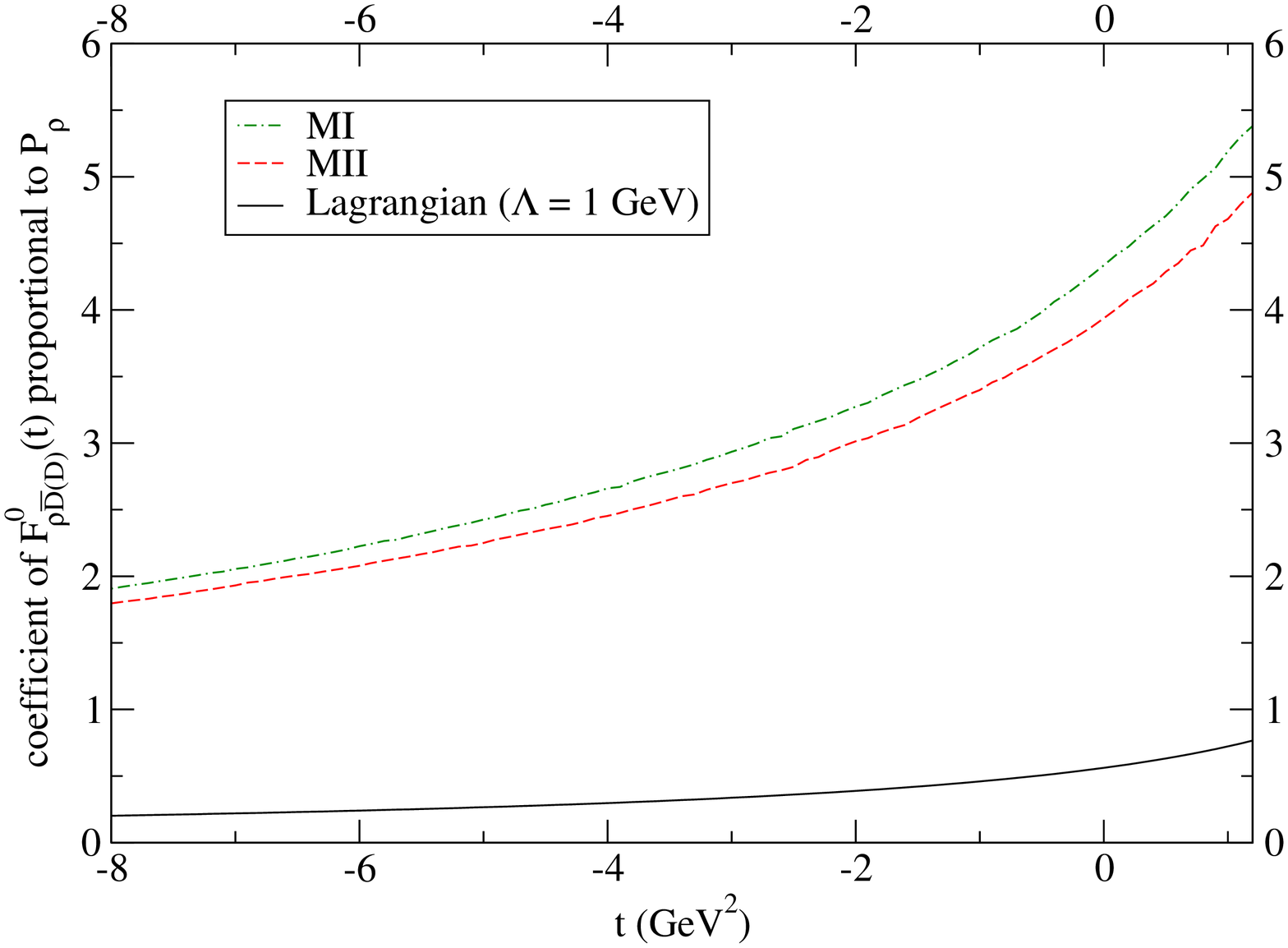}}
\vspace*{-0.5cm}
\hspace*{0cm}
 \\
\multicolumn{2}{c}{
\scalebox{0.326}[0.326]{\includegraphics{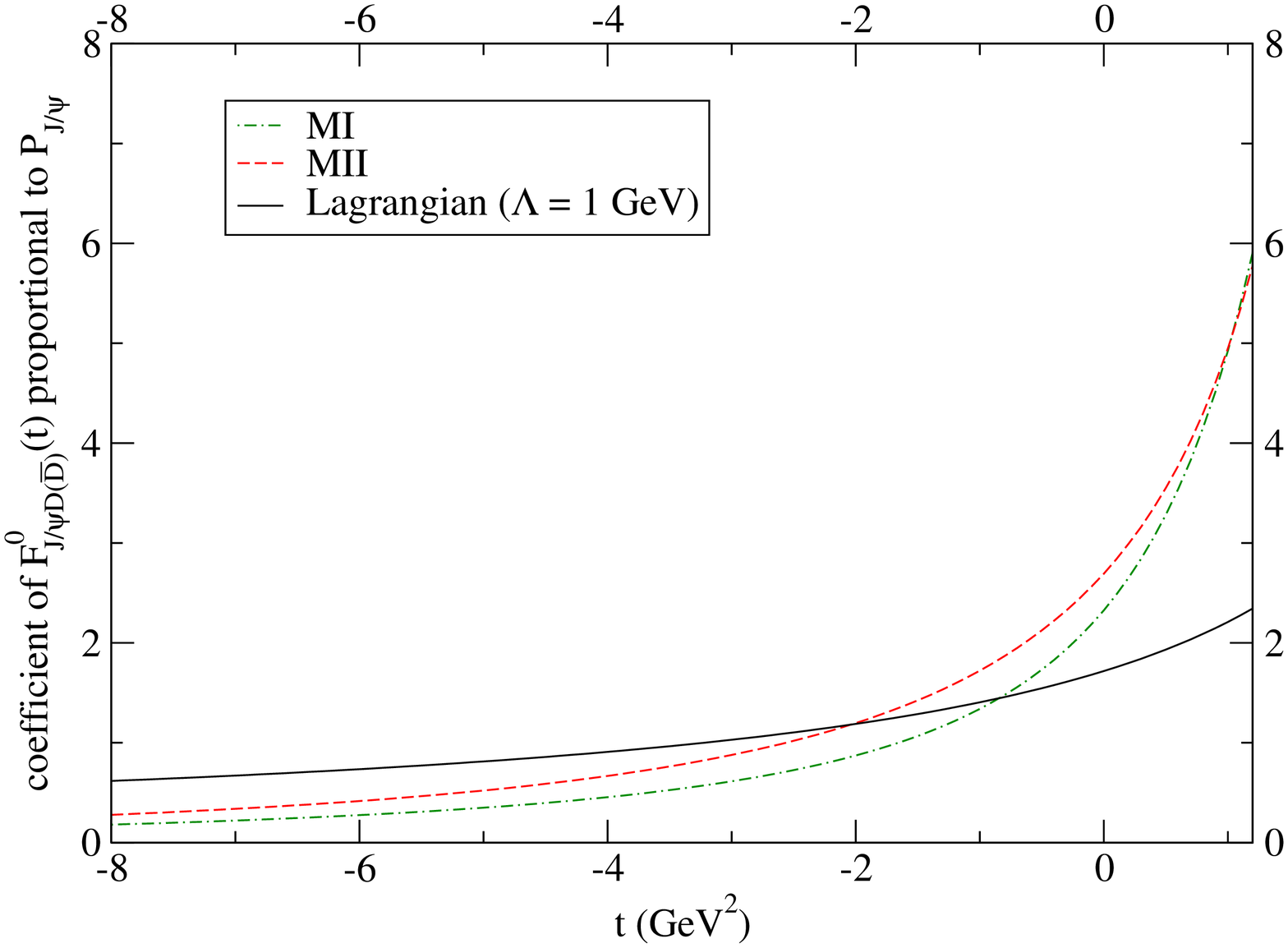}}}
\end{tabular}
\caption[(Color online) Examples of three--point vertices in the non--local NJL model.]{(Color online) Examples of three--point vertices.}
\label{3-vertices}
\end{figure}


Fig.~\ref{3-vertices} shows examples of three--point meson vertices. The curves labeled Lagrangian
are the ad-hoc form factors multiplied by the relevant meson couplings used in Ref.~\cite{bou08}.  Let us first
consider the differences between the Lagrangian and the NJL approaches. We note that a relative agreement only exists
for the coefficient of $F_{\pi \bar D (D^*)}$ proportional to pion momentum, while 
the overall magnitude of the coefficient proportional to the $\rho$--meson momentum 
of $F^{0}_{\rho \bar D (D)}$
 is suppressed and the energy dependence of the 
coefficient proportional to the $J/\psi$ momentum of 
$F^{0}_{J/\psi D (\bar D)}$ is quite different.

\begin{figure}[!htb]
\begin{tabular}{cc}
\hspace*{0cm}
\scalebox{0.326}[0.326]{\includegraphics{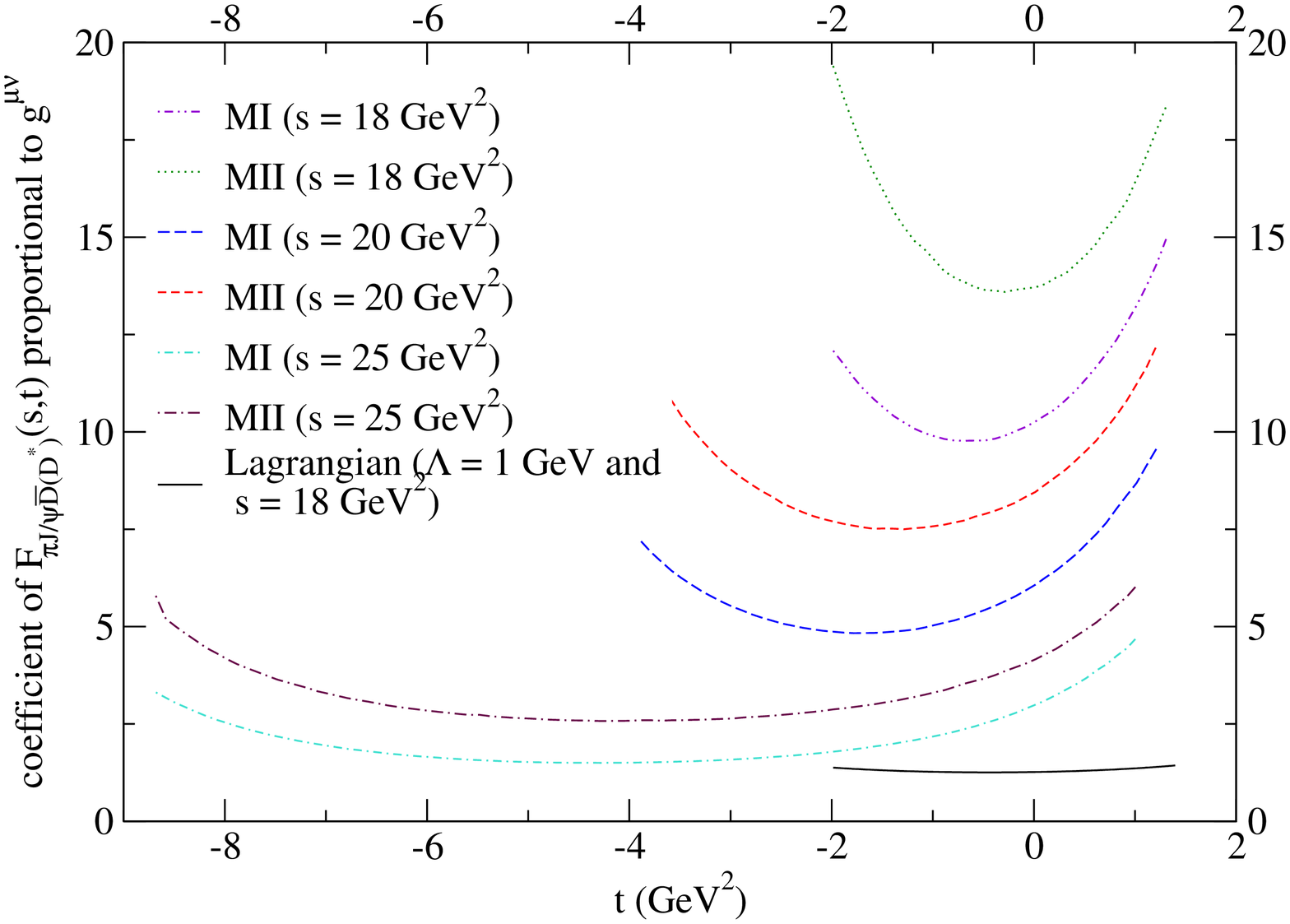}} &
\hspace*{-0.5cm}
\scalebox{0.326}[0.326]{\includegraphics{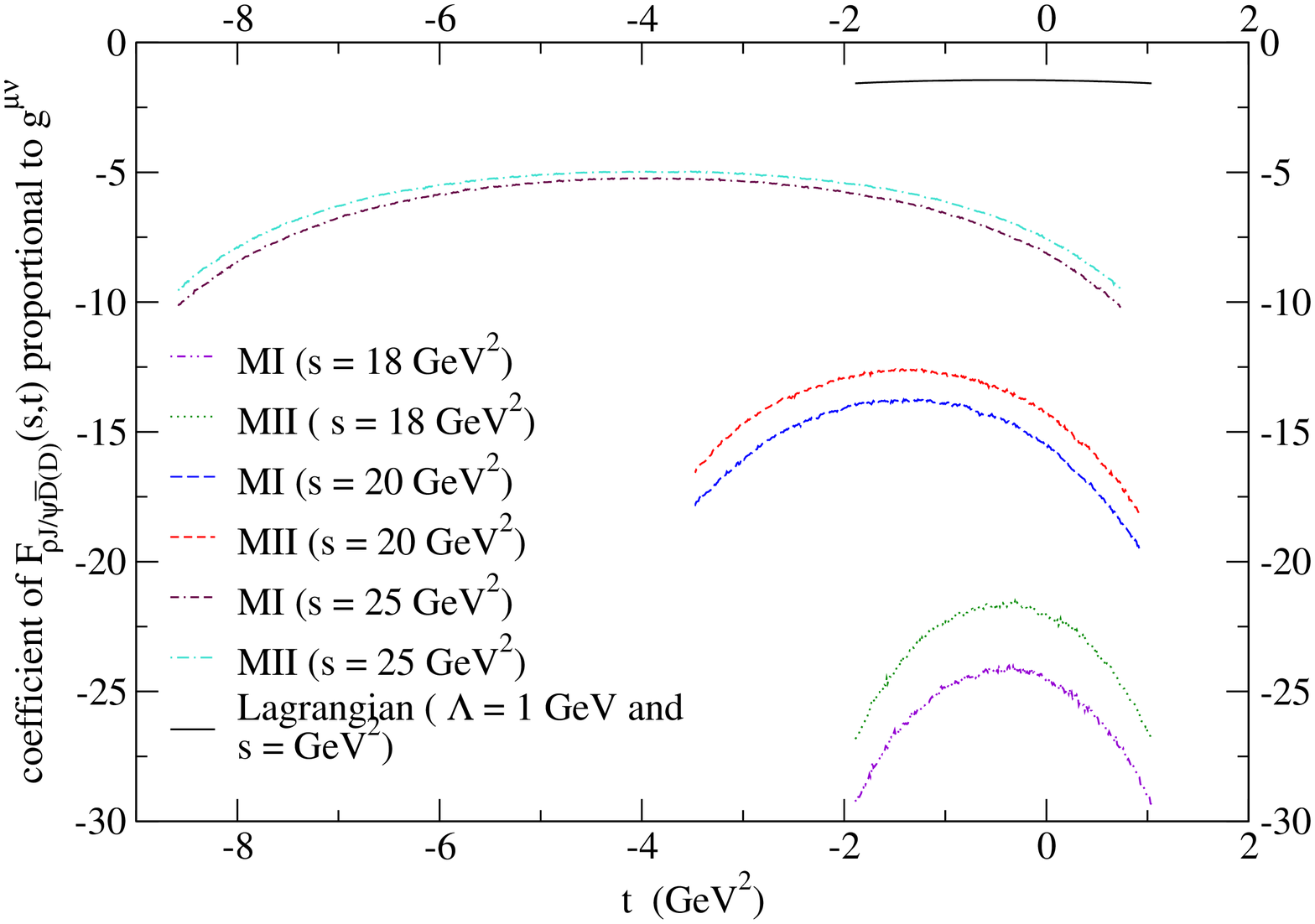}} 
\end{tabular}
\caption[(Color online) Examples of four--point vertices in the non-local NJL model.]{(Color online) Examples of four--point vertices.}
\label{4-vertices}
\end{figure}

Turning to the comparison between MI and MII, we note that the vertices are quite similar both in their overall
magnitude and their energy-dependence with the exception of the ploted coefficient of $F_{\pi \bar D
(D^*)}$. For large
space-like separation, large fluctuations appear for MII. This can be linked to the difficulty of carrying out the principal
value integral  in Eq.~(\ref{JH1H2}) due to the large oscillations in the heavy quark form factor in the loop
integral. As seen in Fig.~\ref{3-vertices}, increasing the number of numerical evaluations reduces the fluctuations.
Although, not apparent in the
figure, this problem exists for all vertices with a light
meson.

Finally, the four--point coefficients of the form factors $F_{\pi J/\psi
\bar D D^*}$ and $F_{\rho J/\psi \bar D D}$ proportional to the metric tensor are
plotted in Fig.~\ref{4-vertices} for three center-of-mass energies.
We note that the form factor used in
the Lagrangian approach of Ref.~\cite{bou08} is very suppressed and quite flat
compared to the NJL equivalent. Moreover, the differences 
between the form factors for MI and MII are slightly more pronounced than observed for the three--point functions 
at least in terms of magnitude.

\begin{figure}[!htb]
\begin{tabular}{cc}
\hspace*{0cm}
\scalebox{0.326}[0.326]{\includegraphics{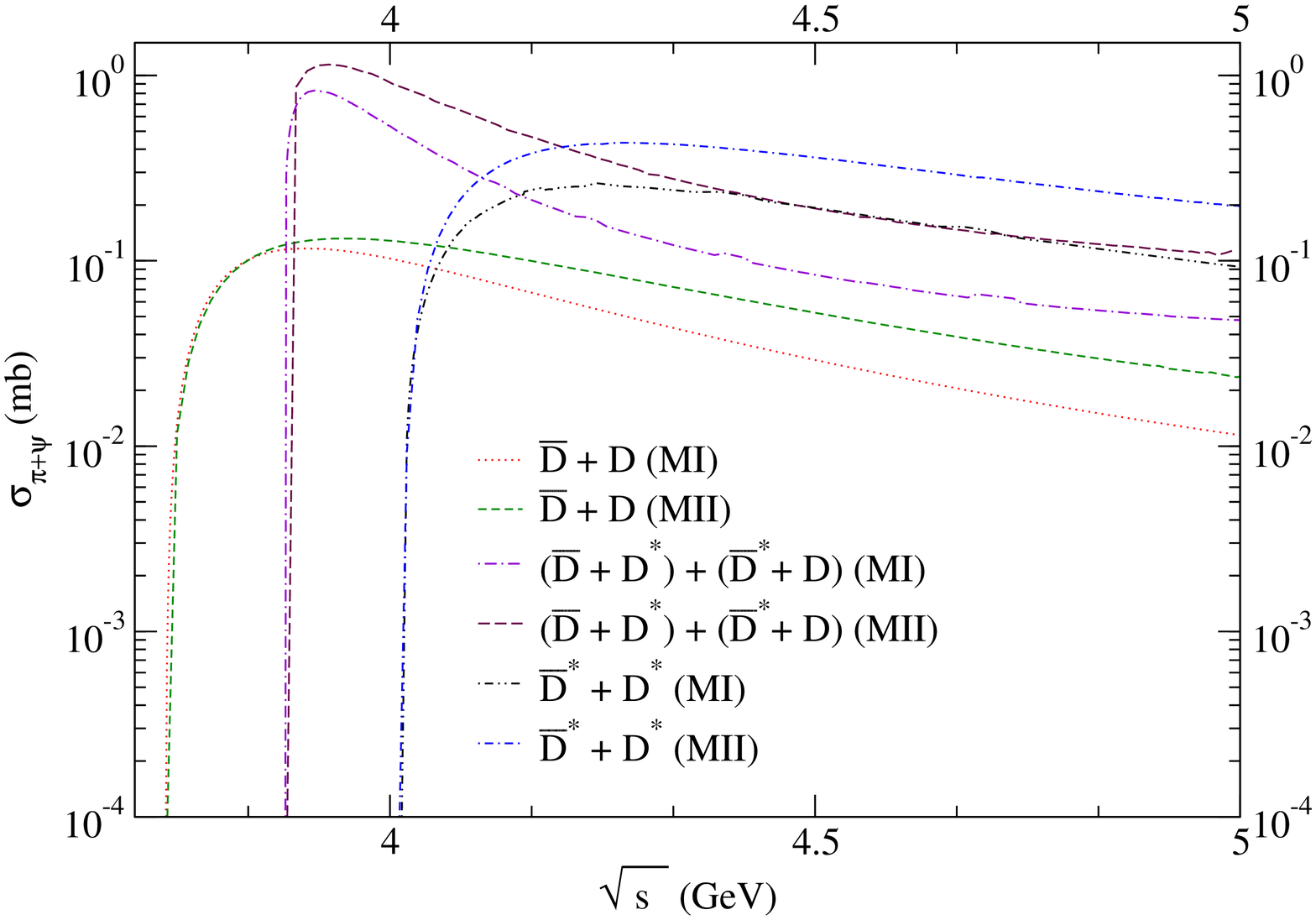}} &
\hspace*{-0.5cm}
\scalebox{0.326}[0.326]{\includegraphics{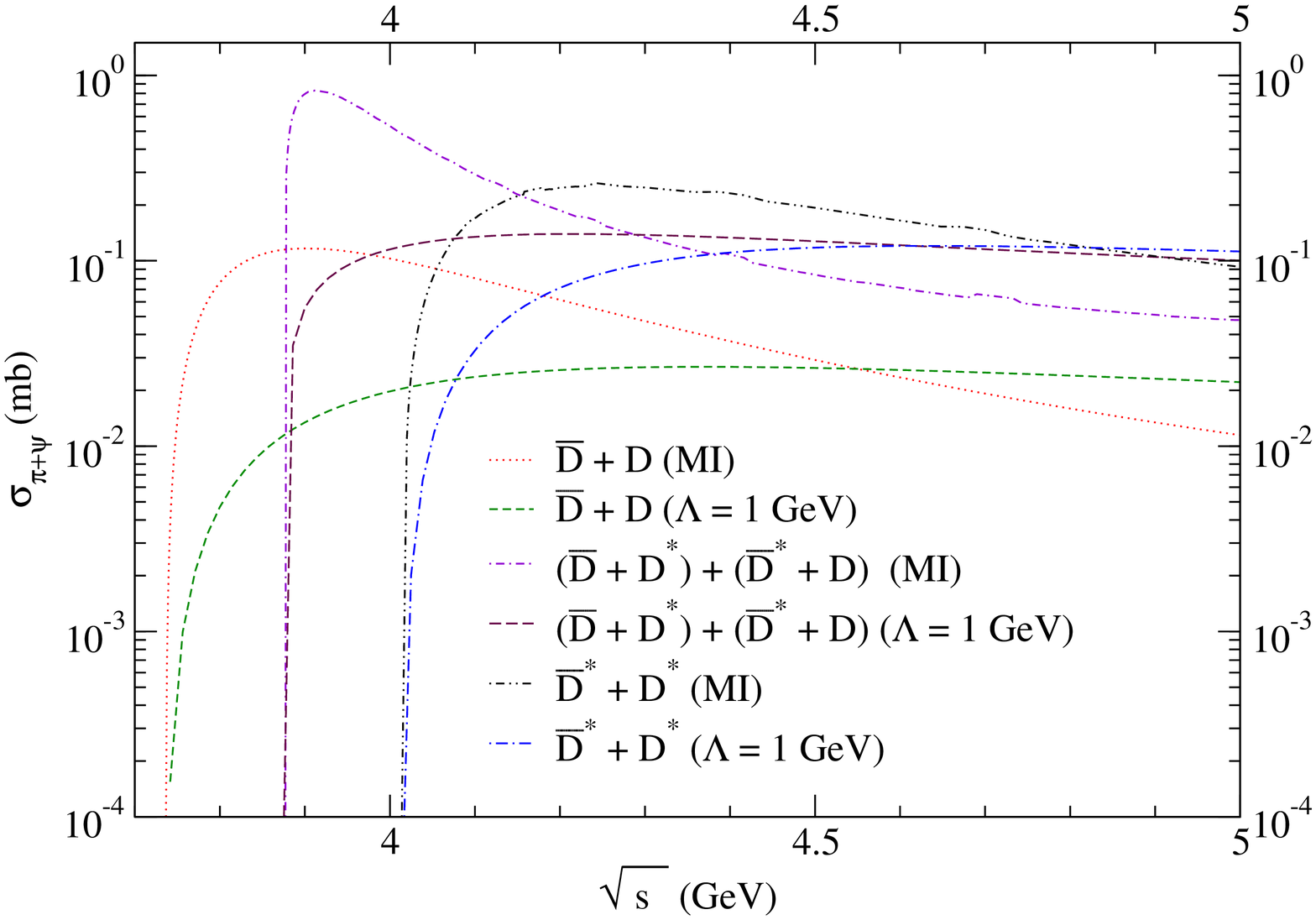}} 
\end{tabular}
\caption[(Color online) cross sections for the $J/\psi--dissociation by a pion in the non-local NJL
model.]{(Color online) cross sections for the $J/\psi$--dissociation by a pion.
Right panel is a comparison between MI and MII, while the left panel is one between MI and the results from
the phenomenological Lagrangian study of Ref.~\cite{bou08} for $\Lambda = 1\,\mbox{GeV}$.}
\label{cs_PIJ_NJL}
\end{figure}

\begin{figure}[!htb]
\begin{tabular}{cc}
\hspace*{0cm}
\scalebox{0.326}[0.326]{\includegraphics{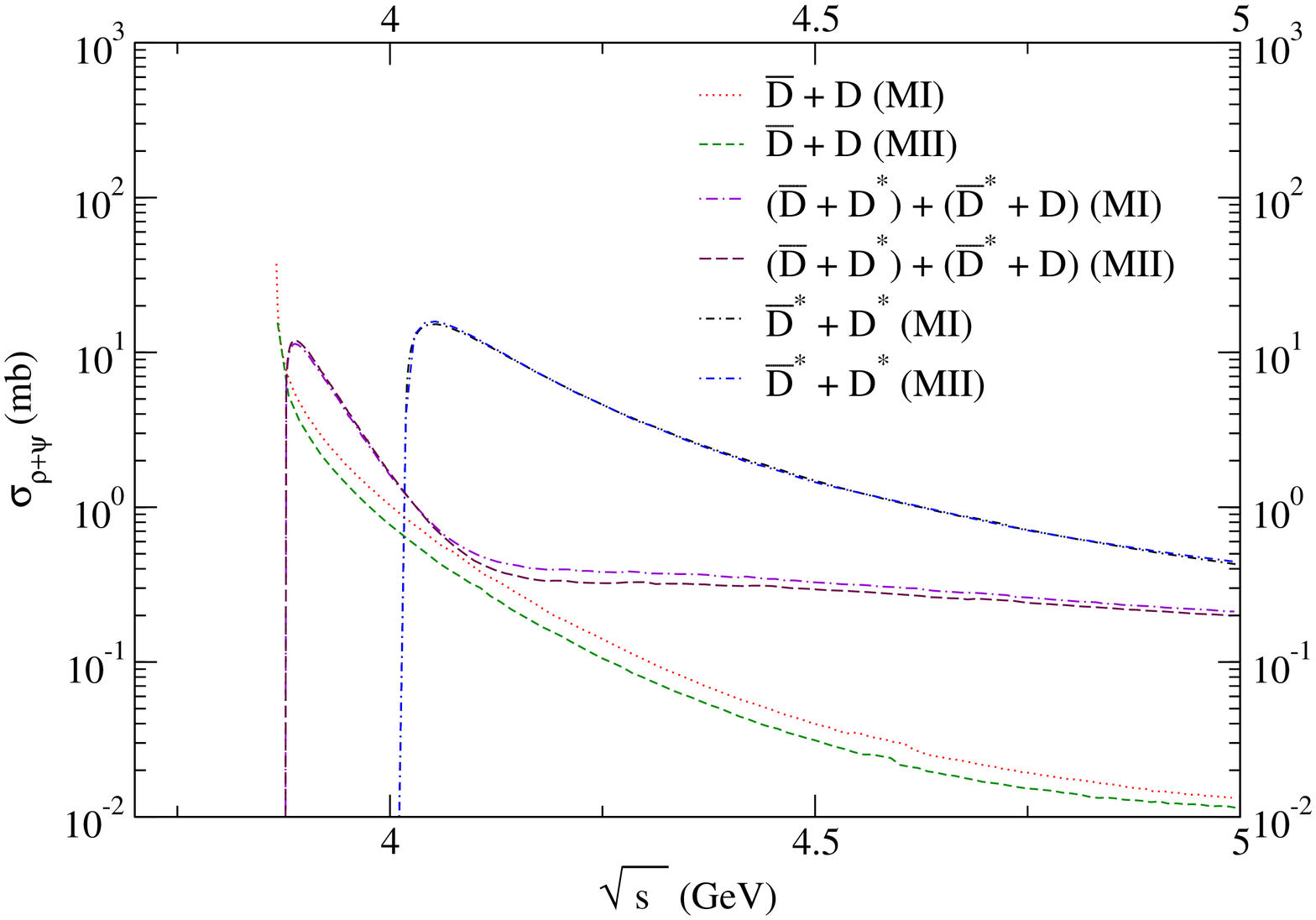}} &
\hspace*{-0.5cm}
\scalebox{0.326}[0.326]{\includegraphics{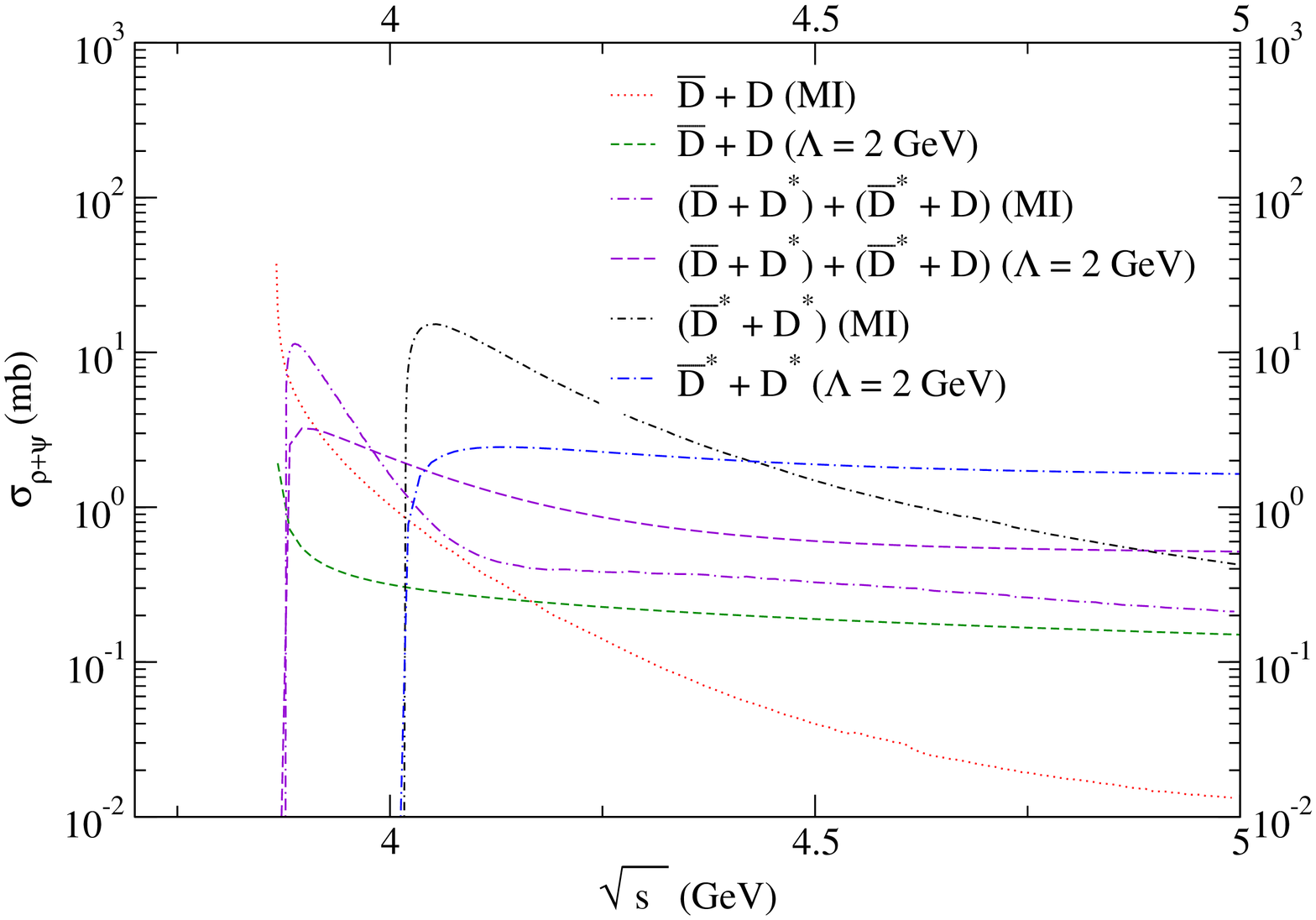}} 
\end{tabular}
\caption[(Color online) cross sections for the $J/\psi$--dissociation by a $\rho$ meson in the non-local NJL
model.]{(Color online) cross sections for the $J/\psi$--dissociation by a $\rho$ meson.
Right panel is a comparison between MI and MII, while the left panel is one between MI and the results from
the phenomenological Lagrangian study of Ref.~\cite{bou08} for $\Lambda = 2\,\mbox{GeV}$.} 
\label{cs_RHOJ_NJL}
\end{figure}

\subsection{Cross sections}

The transition amplitudes can be found in Appendix \ref{amplitudes}. The cross sections 
are plotted in Figs \ref{cs_PIJ_NJL} and \ref{cs_RHOJ_NJL}. 
\footnote{Because of mixing between open charmed
mesons subtle cancellations occur
between sub-amplitudes. For the process $\pi + J/\psi \rightarrow (\bar D + D^*) + (D + \bar D^*)$, for
example, this happens between the $D$-- and $D_1$--exchange channels, and between the
$D^*$-- and $D^*_0$--exchange channels. The requirement that they cancel can be traced back to the tensorial decomposition used
to split the transverse and longitudinal parts of the vector and axial--vector Dirac structures
Eq.~(\ref{tensor}). 
This splitting then induces two $1/q^2$ factors where $q$ is the momentum flowing through the propagator.
One is absorbed in the transverse projector of the vector particle propagator, while the other is further split between the two vertex functions sandwiching
the spin--$0$ propagator. As $q^2 \rightarrow 0 $,
divergences appear. Analytically, when all the sub-amplitudes are summed, they cancel; the splitting being
artificial. However, these cancellations amount to subtracting large numbers. This lead
to a numerical integration problem. Indeed, for certain $\sqrt{s}$, 
the quadrature method employed can require evaluations at points close to $q^2 = 0$. To deal with this problem, we force the cancellations within a
small radius centered around $q^2 = 0$. In some sense, this is an estimation of the
numerical precision associated with the evaluations of the vertices and the propagators. The more precise the evaluations are, the better the cancellation is. A radius value of
$0.05\,\mbox{GeV}^2$ is used here. Relics of the incomplete cancellations usually still remain in
the form of small bumps in the data as seen in Fig.~\ref{cs_PIJ_NJL} around $\sqrt{s} = 4.4\,\mbox{GeV}$ and
$\sqrt{s} = 4.8\,\mbox{GeV}$ for both $\pi + J/\psi \rightarrow (\bar D + D^*) + (D + \bar D^*)$ and $\pi
+ J/\psi \rightarrow \bar D^* + D^*$ ($\pi + J/\psi \rightarrow \bar D + D$ does not have
divergences).}$^,$\footnote{In order to assess the effect of the finite current mass, the cross section for
$\pi + J/\psi \rightarrow \bar D + D$ was re-evaluated with a zero current mass. It was found
that the deviation is small and no greater than $3.2\%$ at $5$ GeV and less than $1.5\%$ near threshold.} 
 We first note that the results for both MI and MII are very similar for
small $\sqrt{s}$ and differ only slightly in magnitude for larger values, and therefore the introduction of two additional parameters in MII is probably not
justified. Comparison with the results of the phenomenological Lagrangian study 
of Ref.~\cite{bou08} shows significant differences with MI from the onset. We note that the maxima of the pion dissociation cross section for $\pi + J/\psi \rightarrow (\bar D + D^*) + (D + \bar D^*)$ and $\pi
+ J/\psi \rightarrow \bar D^* + D^*$ are smaller by about $50\%$ than those found in
the previous non-local NJL study \cite{Iva05} and in the potential model approach \cite{Bar03}.
\footnote{In Refs \cite{Iva05} and \cite{Bar03} specific charged channels are plotted, while here we present isospin
averaged cross sections. In order to make contact with these studies, the $\pi (\rho) + J/\psi \rightarrow
(\bar D + D^*) + (D + \bar D^*)$ cross section has to be divided by two.} Since in the
potential model, the spin-orbit interaction is not modeled the $\pi + J/\psi \rightarrow \bar D + D$
process is not evaluated \cite{Won00}, and a comparison can be made only with Ref.~\cite{Iva05}. Contrary
to the two other pion-induced dissociations the maximum for this process is comparable to that of
Ref.~\cite{Iva05}. 

\begin{figure}[!htb]
\begin{center}
\scalebox{0.5}[0.5]{\includegraphics{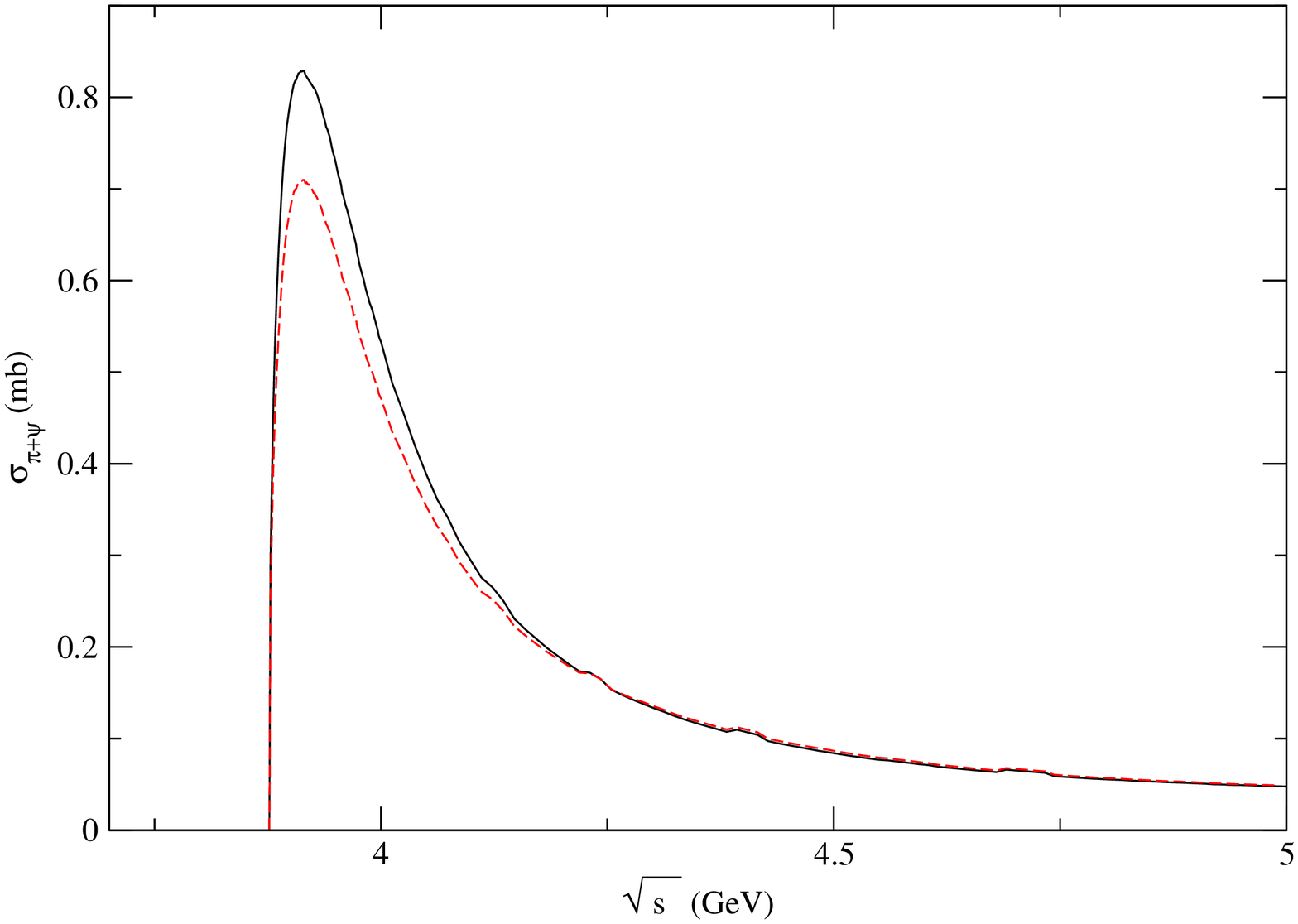}} 
\caption[(Color online) Effect of abnormal parity content on the $\pi + J/\psi \rightarrow (\bar D +D^*) + (\bar D^* + D)$ cross section.]
{(Color online) Effect of abnormal parity content on the $\pi + J/\psi \rightarrow (\bar D
+D^*) + (\bar D^* + D)$ cross section. The dashed and solid lines 
correspond to cross sections without and with the abnormal parity
sub--amplitude.}
\label{csPIJDbDs_parity_comp_NJL}
\end{center}
\end{figure}

A consequence of the mixing between channels is that, it
 is impossible to consider normal parity sub-amplitudes on their own in order to asset the effect of chiral
 symmetry since doing so would entail divergences appearing. However,  this does not prevent us from considering the effect 
 of the abnormal sub--amplitude on 
 $\pi + J/\psi \rightarrow (\bar D + D^*) + (D + \bar D^*)$ process. In Fig.~\ref{csPIJDbDs_parity_comp_NJL}, we observe a
 reduction of the cross section near threshold when the abnormal contribution
 [$\mathcal{M}_{2b}$ of Appendix \ref{amplitudes}] is removed.
Specifically, the maximum ($\sqrt{s} \approx 3.92\,\mbox{GeV}$) is seen to decrease by $14\%$, which is far
less than what was found in phenomenological Lagrangian approach of Ref.~\cite{bou08}.

For the dissociation by the $\rho$ meson, we estimate
 a maximal cross section for $\rho + J/\psi \rightarrow (\bar D + D^*) + (D + \bar D^*)$ comparable 
 to what is found in the potential model \cite{Bar03}, while we find a
maximum for $\rho + J/\psi \rightarrow \bar D^* + D^*$ which is an order of magnitude larger. 
Finally, the trends of the energy behaviors of all $\rho$--induced dissociations are
similar to those found in Ref. \cite{Bar03}.
\section{Conclusion and outlook}
In Ref.~\cite{bou08},  the absolute values of the strength of $\pi$-- and $\rho$--induced dissociations depended
on the choice of form factors and the techniques used to fix their absolute normalizations,
which put into question the robustness of the model. To address this problem, a non-local NJL similar to
that of Refs \cite{Iva05,Pla98} was presented. The vertex form factors
were then calculated from the underlying quark structure, thus reducing some of the uncertainties found in
the previous study of Ref.~\cite{bou08}. We further utilized the fact that in the non-local version, quark form factors can be chosen in
 order to confined analytically the light--quark propagator, i.e., the light quark propagator then has no poles. Doing so
 permitted us to calculate the dissociation cross sections by a $\rho$ meson. However, it was impossible to
 asset the effect of chiral symmetry since mixing between channels prevented the removal
 of the sub-amplitudes where chiral partners are exchanged. This problem did not affect the abnormal parity
 term for $\pi + J/\psi \rightarrow (\bar D + D^*) + (D + \bar D^*)$. Turning it off then
lead to a reduction near threshold as expected. But this decrease was far less substantial than the one observed in the phenomenological
Lagrangian approach of Ref.~\cite{bou08}.

Further work could include calculating the dissociation of higher charmonia as it is relevant within the context
of sequential charmonia melting in the QGP \cite{Kar06}. This would entail improving the heavy--quark
four--point interaction kernels in such a way that they could sustain higher resonances. 
Similarly, other light mesons and, potentially, open strange mesons should be considered. Finally, calculating
semi-leptonic decay constants and other observables, such as the electromagnetic pion form factor, could lead to 
an improvement in the modeling, which would then increase the confidence in the overall magnitude of
$J/\psi$--dissociation cross sections.  

\begin{acknowledgments}
This work was funded in part by the Natural Sciences and Engineering Research
Council of Canada, and by the Fonds Nature et Technologies of Quebec.
\end{acknowledgments}

\appendix
\section{\sf Analytic continuation}
\label{analytic}
In order to calculate the various loop integrals, an analytic continuation to
Euclidean space is needed. In perturbation theory, the usual prescription is the
 Wick rotation, i.e., for the loop momentum: $k^0 \rightarrow ik_4$, which ensures that the
 Feynman boundary condition is encoded and that only the relevant poles are
picked up \cite{Pes95}. For a model with 
analytic confinement, a problem arises. This can be made explicit  by examining an example. 

Consider the loop integral of Eq.~(\ref{fermion_loop}) for the self--energy of a
light meson after a Wick rotation
\begin{eqnarray}
iJ_{ij}(P^2) = -i\int &dk_E& f^2_q\left(k_E-\eta \bar P \right)
f^2_q\left((k_E+(1-\eta)\bar P\right)\nonumber \\
&\times& {\rm Tr}\left[\Gamma_{i} S_q\left(k_E-\eta \bar P \right)
\tilde\Gamma_{j} S_q\left((k_E+(1-\eta)\bar P\right)\right]
\end{eqnarray}
where an arbitrary momentum shift of the loop momentum parametrised by $\eta$ has been
done and $\bar P = \left(-iP^0,\vec{P}\right)$ with $\bar P^2 = -P^2$. Note that the 
squared light-quark form factor has a square-root dependency
since it is defined through Eq.~(\ref{mq}) in the Hartree approximation.
Furthermore, in Euclidean space, the arguments of the form factors and the quark propagators 
are in general complex. The conjunction of these two elements
can then lead to the appearance of branch cuts making the integrand ill-behaved. 

To make this explicit, the chiral limit of the loop integral is taken, namely
\begin{eqnarray}
J(P^2)_{ij} = -\int &dk_E& \sqrt{\frac{
\left[1-e^{-\mu z_{-\eta}}\right]
\left[1-e^{-\mu z_{1-\eta}}\right]e^{-\mu\left(z_{-\eta}z_{1-\eta}\right)}
}{z_{-\eta}z_{1-\eta}}}
g\left(k_E^2,k_E\cdot \bar P, \bar P^2\right) \nonumber \\
\end{eqnarray}
where the arguments are defined as $z_\delta = \left(k_E+\delta \bar P\right)^2$ and the function $g\left(k_E^2,k_E\cdot
\bar P, \bar P^2\right)$ is the product of the Dirac and flavor traces.
An Euclidean coordinate transformation  is then performed 
to go from the Cartesian coordinate system to the spherical coordinate one. This
 is done through \cite{Pes95}
\begin{equation}
k_E = (k_4, \vec{k}) \rightarrow (ky, k \sqrt{1-y^2}\sqrt{1-x^2} \cos \phi,  k
\sqrt{1-y^2}\sqrt{1-x^2}  \sin
\phi, k \sqrt{1-y^2}x).
\end{equation}
Furthermore, to reduce the number of non--trivial integrals to carry out the fermion-loop is
evaluated in the rest--frame of the meson. The complex arguments become
\begin{eqnarray}
z_{-\eta} = k^2 +2i\eta k y P^0 -\eta^2 P^2,
 \quad z_{1-\eta} = k^2 -2i(1-\eta) k y P^0 -(1-\eta)^2 P^2
\end{eqnarray}
with $P = \left(m_M, 0\right)$.
\begin{figure}[!htb]
\begin{center}
\begin{tabular}{cc}
\scalebox{0.4}[0.4]{\includegraphics{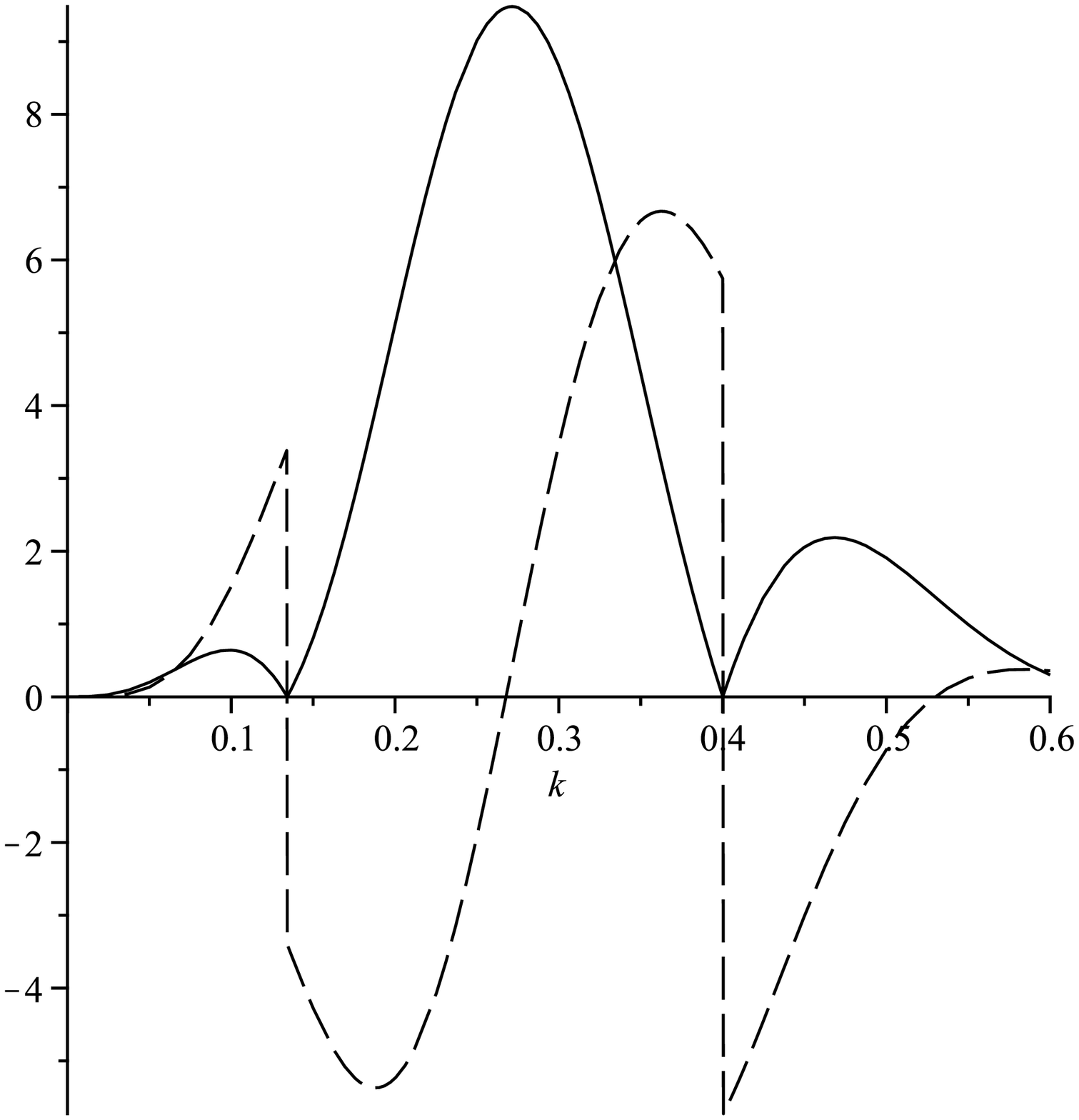}}&
\scalebox{0.4}[0.4]{\includegraphics{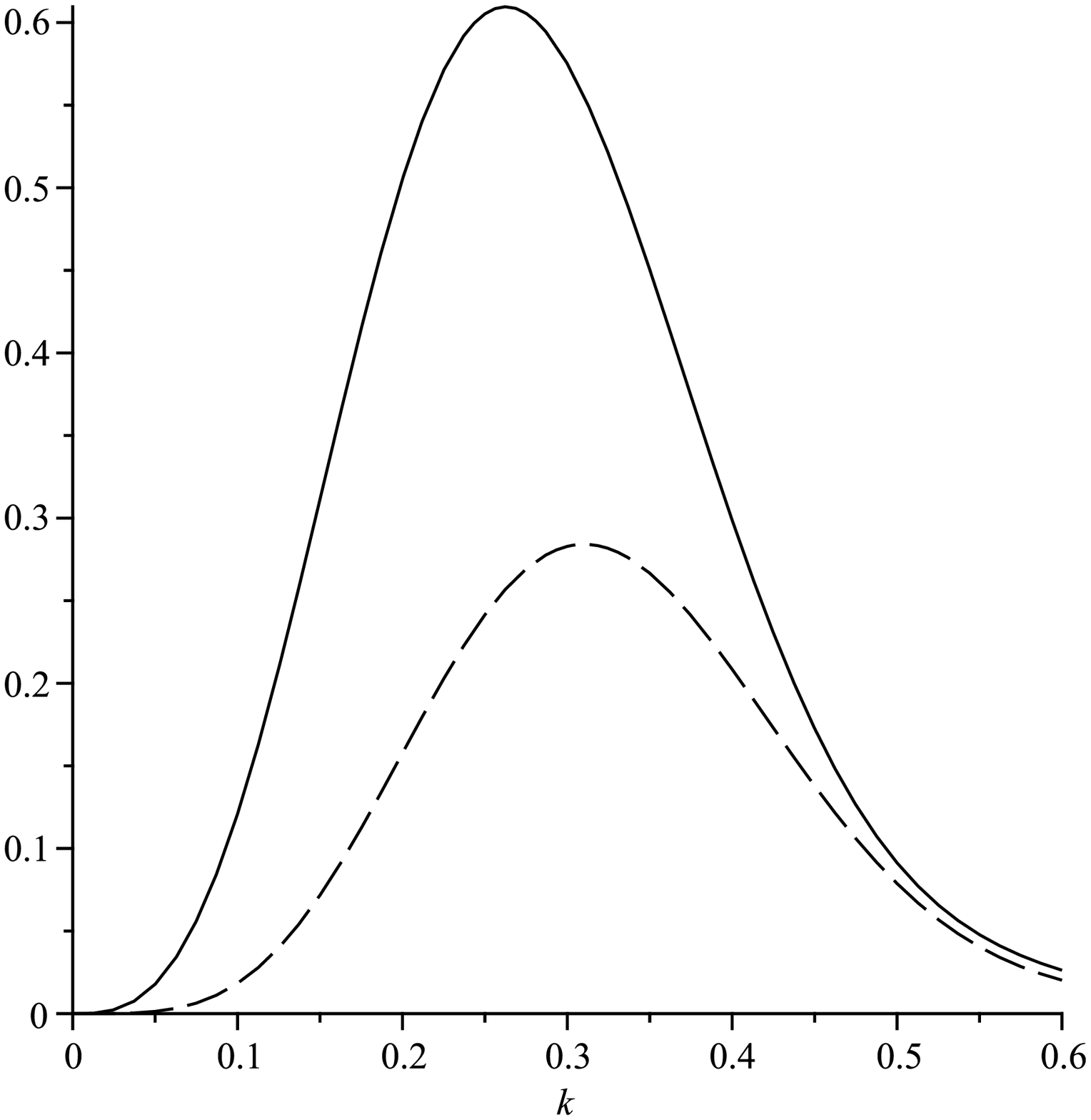}}
\end{tabular}
\caption[Real and imaginary parts of the light-meson self--energy integrand.]
{Real and imaginary parts of the integrand with the trace function omitted. The
solid and dash curves are the real and imaginary parts respectively.  The left
panel is for $\eta = 0.1$ while the right panel is for $\eta=0.45$.}
\label{branch_cuts}
\end{center}
\end{figure}

For the case of $P^2 = m_\rho^2$, both the real and imaginary parts of the integrand with the trace
function $g$ factored out\footnote{An overall minus sign and the constant numerical factor have also been
omitted.} are plotted
as a function of $k$ in Fig.~\ref{branch_cuts} for $y=0.9$ and two values of $\eta$. We note that for
$\eta=0.1$ the imaginary part has several discontinuities which occur when the real part
is zero. 
However, for $\eta=0.45$, both the real and imaginary parts are smooth
functions of $k$. This latter statement turns out to be true for all $y$ in $[-1,1]$. Moreover,
such a behavior can be identified for a range of $\eta$ values. It is then possible to 
evaluate the integral for several $\eta$s within this range and check that the results are equal
as expected from translational invariance. 

We note that by choosing $\eta=\frac{1}{2}$
the arguments of the square-root becomes real and positive for all value of $k$ and $y$. 
In order words, the two squared light-quark form
factors are complex conjugate of each other. It is important to remark that this is the
case only because the evaluation of the fermion--loop is carried in the rest-frame of
the meson. Thus by doing the appropriate shift of momentum, the evaluation of the integral
in the light--meson's rest--frame is numerically tractable. Moreover, this technique can be
applied straight-forwardly to loop integrals of three-- and four--point correlations.
 
However, the above procedure cannot be used  when evaluating
correlation functions with no external light-meson. The simplest example
is for the open charmed self--energies where there is only one squared light-quark form
factor rather than two. The requirement that there are no branch cuts can then be
implemented by shifting the loop momentum in such a way that the
integral reads
\begin{eqnarray}
iJ(P^2)_{ij} = -i\int &dk_E& f^2_Q\left(k_E+\bar P \right)
f^2_q\left(k_E\right) {\rm Tr}\left[\Gamma_{i} S_q\left(k_E\right)
\tilde\Gamma_{j} S_Q\left((k_E+\bar P\right)\right].
\label{loop_open_charm}
\end{eqnarray}
The argument of the squared light-quark form factor is then always positive and real.
However, doing so does leave the possibility that for a certain  $P^2$ 
the Euclidean heavy quark propagator can go on its mass-shell for some of the
$k_E$--integral points. The required continuation prescription can then be found by
going back to the Feynman boundary condition.
\begin{figure}[!htb]
\begin{center}
\scalebox{0.7}[0.7]{\includegraphics{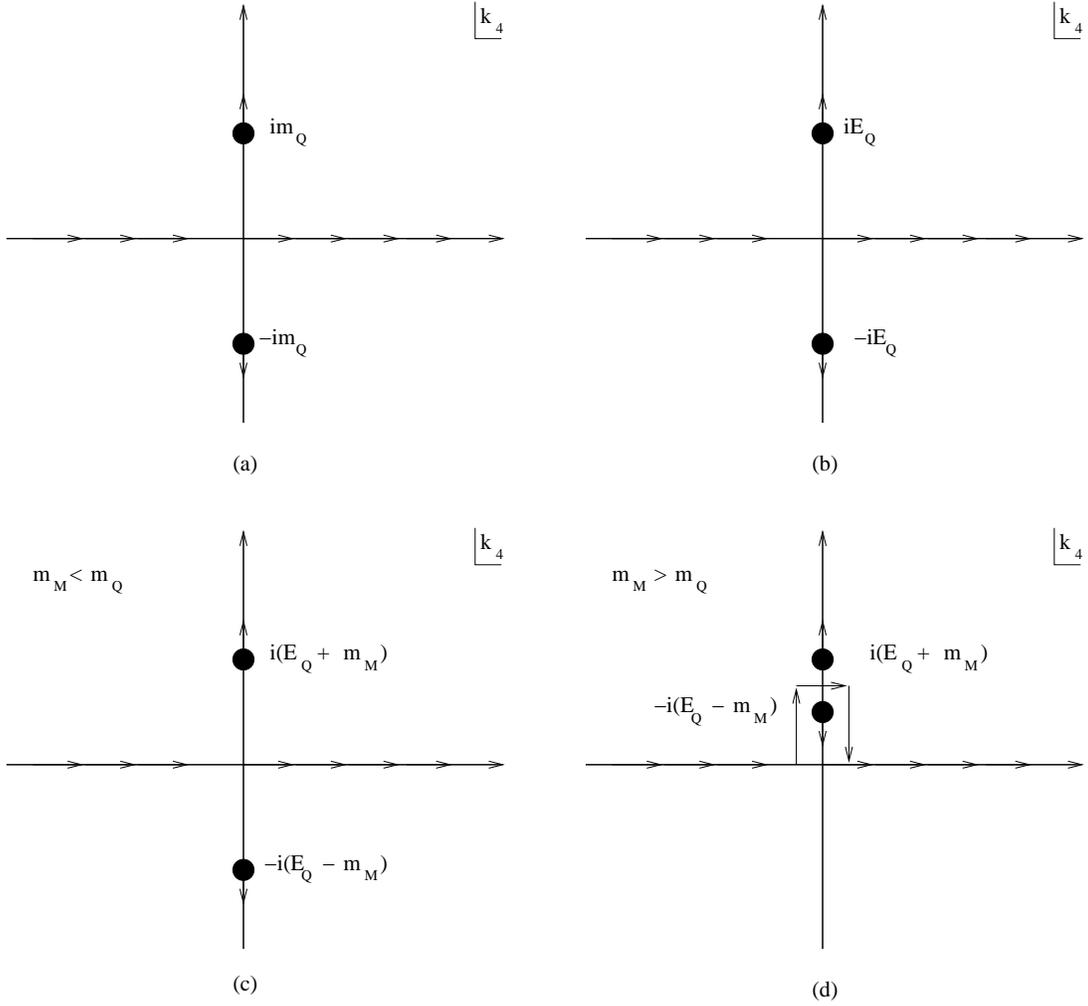}}\\
\caption[Line contours for evaluating the open charmed self--energies.]
{Line contours for evaluating the open charmed self--energies -- (a) $m_H =0$ and
$\left|\vec{k}\right|=0$, (b) $m_H = 0$ and $\left|\vec{k}\right| \neq 0$, 
(c) $m_H < m_Q$ and $\left|\vec{k}\right| \neq 0$, and  $m_H > m_Q$ and 
$\left|\vec{k}\right| \neq 0$. The arrows on the poles indicate in which direction they
moved as  $\left|\vec{k}\right|$ is increased. The Feynman boundary condition implies that all the 
contours are closed in the upper-half of the complex plane (not shown here).}
\label{pole_picking}
\end{center}
\end{figure}
The heavy quark denominator in the loop
integral can be written in Euclidean Cartesian coordinates as
\begin{equation}
(k_E + \bar P) + m_Q^2 = k_4^2 + \left|\vec{k}\right|^2 -2ik_4m_M - m_M^2 +
m_Q^2
\end{equation}
where $m_M$ is the meson mass.
When $\left|\vec{k}\right| = 0$ and $P^2=0$, the poles are at $\pm im_Q$ and the
Feynman prescription dictates that only the positive pole residue should contribute
to the line integral. Keeping the meson mass
to zero, but increasing $\left|\vec{k}\right|$, we note that the poles move
away from the origin. Thus the contour depicted  
in upper-left panel of Fig.~\ref{pole_picking} is equivalent to the Wick 
continuation. Reinstating a meson mass
leads to two cases. The first one is when $m_M < m_Q$. For a null loop
three--momentum, the poles are on each side of the real $k_4$--axis and as 
$\left|\vec{k}\right|$ increases they move away from each other. In this case, there is
again no need to alter the contour. The second case, which is more interesting,
occurs when $m_H> m_Q$. Both poles are in the upper part of the complex plane for
$\left|\vec{k}\right|=0$. One of the pole eventually migrates to the lower half-plane 
as the loop three--momentum is increased. As this pole crosses the real 
$k_4$--axis a jump occurs in the line integral. This is due 
to the fact that by evaluating the line integral along the $k_4$--axis, 
the residue of the pole, which in the limit where $m_M=0$ and $\left|\vec{k}\right|=0$
should not contribute to the integral, is included. The solution is then to deform the
contour as in the lower-right panel of Fig.~\ref{pole_picking} to exclude this pole between $\left|\vec{k}\right|=0$ and
$\left|\vec{k}\right|=\sqrt{m_M^2-m_Q^2}$ where the latter point is found under
the condition $k_4=0$, i.e.,
when the pole is on the real $k_4$--axis and is about to go into the lower half-plane.
 
The above example is one of two possible
scenarios generally encountered. The other one happens when $P \rightarrow -P$ in
Eq.~(\ref{loop_open_charm}). For $m_H > m_Q$ both poles start in the lower half-plane. Thus, 
the contour has to be deformed now to include the pole required by the Feynman boundary condition
for the $\left|\vec{k}\right|$--interval where it is in the lower half-plane. 

This description can be systematically extended to higher-point correlation functions. 
The final integration  prescription in Euclidean space is then, after choosing the appropriate loop momentum flow,
to evaluate the principal value of the line integral and add or subtract the appropriate residues.

\section{Isovector axial Ward identity}
\label{currents}
It is important to verify that the approximation schemes for the quark and meson propagators 
are consistent with each other and do not break chiral symmetry in the chiral limit.
In our model, the divergences of local currents are 
not zero. Rather residual terms due to the non-local interactions are left.  This problem of constructing
a gauge-invariant non-local theory is well studied and we refer the interested reader 
to the Refs \cite{Bow95,Pla98,Bos91} for a complete treatment. In particular, in Ref.~\cite{Pla98} the 
vector current is explicitly  constructed and the related Ward identity is checked. Thus, only the isovector
axial Ward identity has to be ascertained in order to ensure that chiral symmetry is valid within the
approximation context. 

\begin{figure}[!htb]
\begin{center}
\scalebox{0.65}[0.65]{\includegraphics{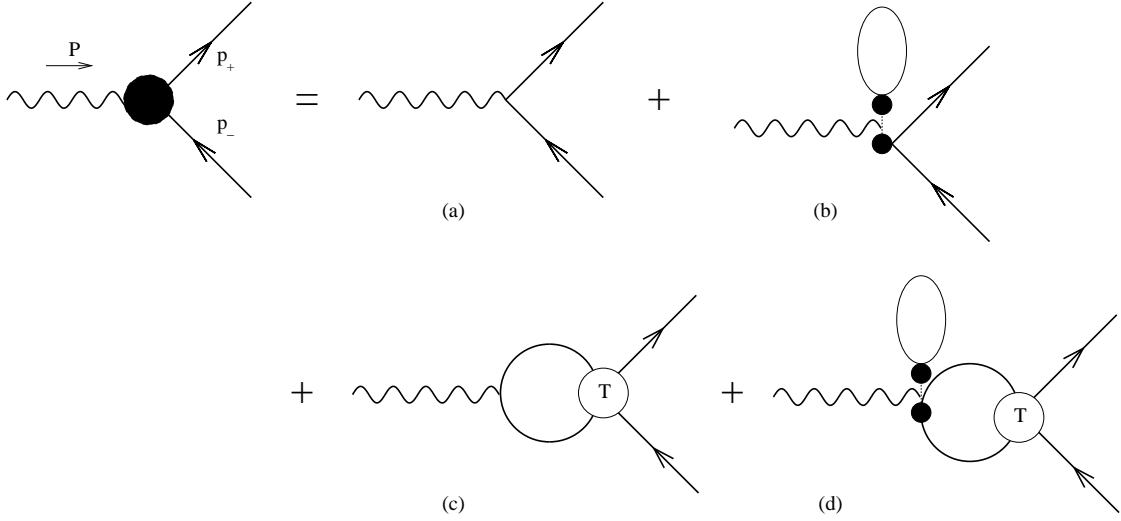}}
\caption[Contributions to the effective isovector axial vertex.]{Contributions to the effective isovector axial vertex. The diagram (a) is the
local contribution while diagrams (b), (c), and (d) are the non-local contributions.} 
\label{axial_ward}
\end{center}
\end{figure}

For the isovector axial symmetry, 
its Ward identity is \cite{Rob94}
\begin{equation}
P_\mu \Gamma_5^{a\mu} = \left\{S^{-1}\left(p_2\right)\gamma_5 +
\gamma_5S^{-1}\left(p_1\right)\right\}\frac{\tau^a}{2}.
\end{equation}
where the momentum flows are given in Fig.~\ref{axial_ward}.
\footnote{From Eq. (\ref{axial_ward}), it is expected that for a finite current mass: 
$P_\mu \Gamma_5^{a\mu} = 2m^q_c.\gamma_5\frac{\tau^a}{2}$.}
Two cases are considered. 
The first one is when there are only scalar and pseudo-scalar four-quark couplings, while 
the second one includes mixing due to the introduction of a vector and axial channels.
In both cases, the term due to the current quark mass is omitted. For the first case, 
they are four different contributions to the isovector axial vertex. These are depicted in
Fig.~\ref{axial_ward}. The first non-local contribution to the divergence (diagram labeled b in
Fig.~\ref{axial_ward}) is due to a scalar fermion-loop
which can be inferred from Eq. (13) of Ref.~\cite{Pla98}. 
Summing the local and first non-local terms yields 
\begin{eqnarray}
P_\mu \tilde \Gamma^{a\mu}_5 &=& P_\mu \gamma^\mu\gamma_5\frac{\tau^a}{2} + P_\mu
J^{5a\mu}_S(P) \nonumber \\
&=& \left\{S_q^{-1}(p_2)\gamma_5 + \gamma_5 S^{-1}_q(p_1)+ if_q(p_1)f_q(p_2)\gamma_5 I_S(P)\right\}\frac{\tau^a}{2} 
\label{vertex_1}
\end{eqnarray}
where the scalar fermion-loop is defined as
\begin{equation}
I_S(P) = G_S\int dk f_q(k)
{\rm Tr}\left[S_q(k)\right]\left(f_q(k+P)+f_q(k-P)\right).
\end{equation}

Next, the contributions due to the pionic resonance in the absence of mixing [diagrams
(c) and (d) of Fig.~\ref{axial_ward}] can be cast as
\begin{eqnarray}
\Gamma^{5b}_{PS} &=& -i\frac{G_P f_q(p_1)f_q(p_2)}{1-G_PJ_{PP}(P)} 
\gamma_5 \tau^a \int dk f_q(k_+)f_q(k_-) {\rm Tr}\left[\gamma_5\tau^a S_q(k_+)P_\mu\tilde
\Gamma^{b\mu}_5S_q(k_-)\right].
\end{eqnarray}
Inserting Eq.~(\ref{vertex_1}) into the above equation  yields the final expression for 
\begin{eqnarray}
\Gamma^{5b}_{PS} &=&  -i\frac{f_q(p_1)f_q(p_2)}{1-G_PJ_{PP}(P)} 
\gamma_5 \frac{\tau^b}{2} \left[1-G_PJ_{PP}(P)\right]I_S(P) \nonumber \\
&=& -if_q(p_1)f_q(p_2) \gamma_5 \frac{\tau^b}{2}I_S(P).  
\end{eqnarray}
Summing this contribution and that of Eq.~(\ref{vertex_1}) verifies the axial Ward
identity.

Adding the vector and axial channels leads to an additional contribution to
Eq.~(\ref{vertex_1})
due to the vector insertion in the fermion-loop. The divergence of the resulting
non-local current in momentum-space is inferred from Eq.(10) of Ref.~\cite{Pla98} and
reads
\begin{eqnarray}
P_\mu J^{5a\mu}_V(P) &=& if_q(p_1)f_q(p_2) I_V(P)\frac{\displaystyle{\not}
P}{\sqrt{P^2}}\gamma_5\frac{\tau^a}{2} 
\label{vertex_V}
\end{eqnarray}
where the vector fermion-loop is
\begin{equation}
I_V(P) = G_V\int dk {\rm Tr}\left[\frac{\displaystyle{\not}
P}{\sqrt{P^2}}S_q(k)\right]f_q(k)\left(f_q(k+P)-f_q(k-P)\right).
\end{equation}
The contribution due to the pion intermediate state is then 
\begin{eqnarray}
\Gamma^{5b}_{PS} &=&  -if_q(p_1)f_q(p_2)\left\{I_S(P)+\frac{\displaystyle{\not}
P}{\sqrt{P^2}}I_V(P)\right\}\gamma_5\frac{\tau^b}{2} 
\end{eqnarray}
which again cancels both the second term of Eq.~(\ref{vertex_1}) and Eq.~(\ref{vertex_V})
thus verifying the axial Ward identity for this extension.
\section{\label{Decays}Decays}

\subsection{$g_{\pi D^* D}$ coupling}
For the dissociation processes studied, all three--point vertices are evaluated with one 
external particle off-shell. Moreover,  for most of them the kinematics do not allow to have all
three mesons on-shell. One exception is for the  $D^* \rightarrow D + \pi$ decay process which has been
measured experimentally by CLEO \cite{Ana03} 

We then wish to use this experimental information to
constrain the parameter set. To do so, the expression of the decay width as a
function of the on-shell three--point coupling is written down:
\begin{equation}
\Gamma_{D^{*+}\rightarrow D^+ + \pi^0} = \frac{g^2_{\pi D^* D}\left|\vec{P}_\pi\right|^3}{48 m_{D^*}^2}
\end{equation}
where $\left|\vec{P}_\pi\right|$ is the centre-of-mass three-momentum. The second step then consists
in relating the coupling to the the extended NJL model. The associated meson form factor can be decomposed
into
\begin{equation}
F^{\mu}_{D^{*+} \rightarrow D^+ + \pi^0} = F_0 P_\pi^\mu + F_1P_{D}^\mu,
\end{equation}
with the coupling given by
\begin{equation}
g_{\pi D^* D} = \sqrt{2}\left(F_0 -F_1\right)
\label{g_piDsD}
\end{equation}
where four-momentum conservation $P_{D^*} = P_\pi + P_D$ and orthogonality $\epsilon\left(P_{D^*}\right)\cdot
P_{D^*} = 0$ have been used, and the factor $\sqrt{2}$ is to account for isospin (
 the coupling $g_{\pi D^* D}$ being defined to be equal to $g_{D^{*+}\rightarrow D^0+\pi^+}$
 \cite{Ana03}).
\subsection{$J/\psi$ decay into a dilepton}
\begin{figure}[!htb]
\begin{center}
\scalebox{0.65}[0.65]{\includegraphics{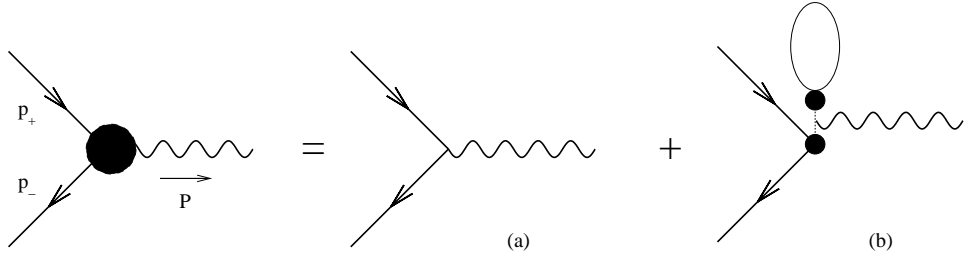}}
\caption[Contributions to the effective quark-photon vertex for the decay of the $J/\psi$ into a dilepton.]
{Contributions to the effective quark-photon vertex for the decay of the $J/\psi$
into a dilepton. Diagram (a) and (b) are the local and non-local contributions
respectively. The latter is due to a fermion-loop with a vector insertion.}
\label{EM_vertex}
\end{center}
\end{figure}

We now turn to the calculation of the
decay of the $J/\psi$ into a dilepton. Fig.~\ref{EM_vertex} shows the
contributions to the effective quark-photon vertex. It is assumed that the dominant 
behavior will be given by diagrams at leading order in $1/N_C$. Thus, 
only the transition due to constituent quark loop in the direct channel 
will be considered; exchange diagrams and mesonic fluctuations are ignored. 
It is important to note that this approximation is consistent with the Ward 
identity \cite{Pla98}.
The transition amplitude of an on-shell $J/\psi$ into a photon is then 
\begin{eqnarray}
i\mathcal{M}^{\mu\nu}_{\psi\gamma}(P) &=& -\int
dk {\rm Tr}\left[i\chi^\mu(k,P)iS_Q(k_-)i\Gamma^\nu_{EM}(k,P)iS_Q(k_+)\right] \nonumber
\\ 
&=& -g_\psi e QT^\mu_\alpha\int dk f_Q(k_+)f_Q(k_-)
{\rm Tr}\left[\gamma^\alpha S_Q(k_-)\Gamma^\nu_{EM}(k,P)S_Q(k_+)\right] \nonumber \\
&=& -g_\psi e QT^{\mu \nu} \left[I^{L}_T(P)+I^{NL}_T(P)\right]
\end{eqnarray}
where $Q$ is the electric charge of the heavy quark, $e = |e|$, 
the minus sign in front of the integral is due to the fermion-loop, and the wavefunction meson label is
suppressed. The coupling constant between the full current and the photon field, $A$, is derived
from
\begin{equation}
\mathcal{L}_{EM} = -eQJ_{EM}^\mu(x) A_\mu(x).
\end{equation}
And coupling the photon to a dilepton finally yields the decay width
\begin{equation}
\Gamma_{\psi \rightarrow l^+l^-} = \frac{e^2}{6\pi}
\left[\frac{\mathcal{M}_{\psi\gamma}^{2}}{m_\psi^4}\right]
\frac{|\vec{P}|^3}{m_\psi^2}
\label{gamma_J_epem}
\end{equation}
where $|\vec{P}|^2 = \frac{m_\psi^2}{4}-m_l^2$.

The main ingredient left to specify is thus the effective electromagnetic vertex.
In the heavy quark sector, only a vector coupling is introduced. Thus there is
only one correction term to the local vertex and we can write
\begin{equation}
\Gamma^\mu_{EM}(p,P) = \gamma^\mu + \Gamma^\mu_{V}(p,P)
\end{equation}
where
\begin{eqnarray}
\Gamma^\mu_{V}(p,P) &=&  f_Q(p_+)f_Q(p_-)\gamma^\nu \left[iG_V I^\mu_\nu(P)\right]
\end{eqnarray}
and 
\begin{eqnarray}
I^\mu_\nu(P) &=&  -\int_0^1 d\lambda \int dk \,{\rm Tr}\left[\gamma_\nu
S_Q(k)\right] \frac{\partial}{\partial k^{\mu}} f_Q(k+\lambda P)f_Q(k-(1-\lambda)P).
\end{eqnarray}
Inserting the electromagnetic vertex into the transition amplitude gives
\begin{eqnarray}
i\mathcal{M}^{L;\mu\nu}_{\psi\gamma}(P) &=& - g_\psi e Q T^\mu_\alpha \int dk f_Q(k_+)f_Q(k_-)
{\rm Tr}\left[\gamma^\alpha S_Q(k_-)\gamma^\nu S_Q(k_+)\right] \nonumber \\
&=& -g_\psi e Q T^{\mu\nu} I^{L}_T
\end{eqnarray}
and
\begin{eqnarray}
i\mathcal{M}^{NL;\mu\nu}_{\psi\gamma}(P) &=& - g_\psi e Q T^{\mu\beta} 
\left[G_V J^T_\psi(P^2)\right]I_{\beta}^{\nu}(P) = -g_\psi e Q T^{\mu\nu} I^{NL}_T 
\end{eqnarray}
where, for the non-local term, Eqs.~(\ref{fermion_loop}) and
(\ref{pole}) have been used and the scalar integrals are defined as
\begin{eqnarray}
I^{L}_T &=& 4N_C\int dk\, f_Q(k_+)f_Q(k_-)\frac{\left(m^{2}_Q +\frac{P^2}{4} -\frac{k^2}{3} 
-\frac{2}{3}\frac{\left(k\cdot
P\right)^2}{P^2}\right)}{\left[k_+^2-m^{2}_Q\right]\left[k_-^2-m^{2}_Q\right]}
\end{eqnarray}
and 
\begin{eqnarray} 
I^{NL}_T &=& \frac{8N_C}{3}\int_0^1 d\lambda \int dk
\frac{\left(k^2-\frac{\left(k\cdot
P\right)^2}{P^2}\right)}{k^2-m^{2}_Q}
\frac{d}{dk^2}  f_Q(k+\lambda P)f_Q(k-(1-\lambda)P).
\end{eqnarray}
From the non-local scalar integrals, we note that the final result will dependent on the
interpolation path. This is due to the fact that the coupling between the photon and the
$J/\psi$ is transverse and, thus, not constrained by the underlying
current conservation.

\subsection{Pion decay constant}
The leptonic decay of the pion into a muon and a muonic anti-neutrino can be studied by considering 
the coupling of the pion-field to the
isovector axial current \cite{Mos99}. Formally, the coupling is inferred
from the matrix element
\begin{equation}
\int dx e^{iP\cdot x} \left<0\right|T\left(\pi^a(x)J_5^{b\mu}(0)\right)\left|0\right>
\end{equation}
where translational invariance has been invoked and the pion momentum is outgoing.
Near the pion pole, it becomes \cite{Pes95}
\begin{equation}
\int dx e^{iP\cdot x} \left<0\right|T(\pi^a(x)J_5^{b\mu}(0)\left|0\right> \rightarrow
\frac{ i\mathcal{M}^{\mu}_{AP}
\delta^{ab}}{P^2-m_\pi^2}
\end{equation}
where the transition amplitude $\mathcal{M}^{\mu}_{AP}$ is given by
\begin{equation}
i\mathcal{M}^{\mu}_{AP}P_\mu \chi^a(p,P) =
\left.\left[i\Gamma^{5a}_{PS}\left(P^2-m_\pi^2\right)\right] \right|_{P^2\approx
m_\pi^2}.
\end{equation}
The latter is a consequence of the isovector axial current being dominated by the
pion-resonance contribution near the pole \cite{Rob94}.
Moreover, for leptonic decay, the transition amplitude is usually parametrised as
\begin{equation}
i\mathcal{M}^{\mu}_{AP} = if_\pi P^\mu.
\end{equation}
Putting everything together leads to the expression
\begin{equation}
\left. \left[i\Gamma^{5a}_{PS}\left(P^2-m_\pi^2\right)\right] \right|_{P^2\approx
m_\pi^2} = if_\pi m_\pi^2 \chi^a(p,P).
\label{f_pi}
\end{equation}

The evaluation of the pion decay constant is thus reduced to that 
of the pseudo-scalar contribution near the pion pole. Consider first a theory without any
vector or axial four-quark couplings. The pseudo-sclalar contribution near the pion pole is given at leading $N_C$ order by 
\begin{eqnarray}
\Gamma^{5b}_{PS} &\approx& \frac{\chi^a(p,P)}{P^2-m_\pi^2} \frac{g_\pi}{2}
\Bigg \{ \left(1-G_P J_{PP}(P)\right)\delta^{ab}I_S(P)  \nonumber \\
&+&2m^q_c\int dk f_q(k_+)f(k_-)
{\rm Tr}\left[\gamma_5\tau^aS_q(k_+)\gamma_5\tau^b S_q(k_-)\right]\Bigg\}
\end{eqnarray}
where the second term is due to finite current quark mass. Using
the definition of the pion wavefunction, the pion decay constant
is extracted from Eq.~(\ref{f_pi}) and reads
\begin{eqnarray}
f_\pi m_\pi^2 \delta^{ab} &=&  -m^q_c\int dk
{\rm Tr}\left[\bar \chi^a(k,P)S_q(k_+)i\gamma_5\tau^b S_q(k_-)\right]  \nonumber \\
&+& \frac{g_\pi}{2}\left(1-G_P J_{PP}(P)\right)\delta^{ab}I_S(P).
\end{eqnarray}
where at the pole the second term is zero. Note also, that because the
quark-pion coupling scales like $1/\sqrt{N_c}$, the pion decay constant will have a 
$\sqrt{N_c}$ dependence.

Introducing vector and axial couplings, the expression for the pion decay constant
then becomes
\begin{eqnarray}
f_\pi m_\pi^2 \delta^{ab}&=&  -m^q_c\int dk
{\rm Tr}\left[\bar \chi^a(k,P)S_q(k_+)i\gamma_5\tau^b S_q(k_-)\right]   \nonumber \\
&+& \frac{g_\pi}{2}\Delta_\pi(P)\left\{I_S(P)+\frac{\displaystyle{\not}
P}{\sqrt{P^2}}I_V(P)\right\} \delta^{ab}.
\label{f_pi_2}
\end{eqnarray}
Again, only the first term survives at $P^2=m_\pi^2$. 

Contrary to the electromagnetic decay of the $J/\psi$ into a dilepton,
the pion decay constant does not dependent on the path. This is due to the fact that the
pion couples to the divergence of the isovector axial current, i.e., its longitudinal
part, which is entirely determined by the axial Ward identity.

\section{\sf Scattering amplitudes for the non-local NJL model}
\label{amplitudes}

\subsection{$J/\psi+\pi \rightarrow \bar D + D$}
\noindent
\begin{eqnarray}
\mathcal{M}^{\rho}_{1a} &=& -F^\alpha_{\pi D (\bar D^*)}(t) \mathcal{D}_{\alpha
\beta}^{D^*}(p_\pi-p_D)
F^{\beta \rho}_{\psi \bar D (D^*)}(t), \\ 
\mathcal{M}^{\rho}_{1b} &=& -F^\alpha_{\pi \bar D (D^*)}(u) \mathcal{D}_{\alpha
\beta}^{D^*}(p_\pi-p_{\bar D})
F^{\beta \rho}_{\psi D (\bar D^*)}(u), \\ 
\mathcal{M}^{\rho}_{1c} &=& F^\rho_{\pi \psi \bar D D}(s,t)
\end{eqnarray}
where $t=\left(p_\pi - p_D\right)^2$ and $u=\left(p_\pi - p_{\bar D}\right)^2$. 

\subsection{$J/\psi+\pi \rightarrow \bar D + D^*$}
\begin{eqnarray}
\mathcal{M}^{\mu \rho}_{2a} &=& \sum_{i} F^{\mu;i}_{\pi D^* (\bar D)}(t)
\mathcal{D}_{ij}^{D}(p_\pi-p_{D^*}) F^{\rho;j}_{\psi \bar D (D)}(t), \\ 
\mathcal{M}^{\mu\rho}_{2b} &=&  F^{\mu \alpha}_{\pi  D^* (\bar D^*)}(t) 
\mathcal{D}_{\alpha\beta}^{D^*}(p_\pi-p_{D^*})F^{\beta \rho}_{\psi \bar D (D^*)}(t), \\ 
\mathcal{M}^{\mu\rho}_{2c} &=& F^{\alpha}_{\pi  \bar D (D^*)}(u) 
\mathcal{D}_{\alpha\beta}^{D^*}(p_\pi - p_{\bar D})F^{\beta \mu \rho}_{\psi D^* (\bar D^*)}(u), \\ 
\mathcal{M}^{\mu\rho}_{2d} &=& F^{\mu\rho}_{\pi \psi \bar D D^*}(s,t), \\
\mathcal{M}^{\mu\rho}_{2e} &=& F^{\mu \alpha}_{\pi  D^* (\bar D_1)}(t) 
\mathcal{D}_{\alpha\beta}^{D_1}(p_\pi-p_{D^*})F^{\beta \rho}_{\psi \bar D
 (D_1)}(t), \\ 
\mathcal{M}^{\mu \rho}_{2f} &=& \sum_{i} F^{i}_{\pi \bar D ( D^*_0)}(u)
\mathcal{D}_{ij}^{D^*_0}(p_\pi - p_{\bar D}) F^{\mu \rho;j}_{\psi  D^* (\bar D^*_0)}(u)  
\end{eqnarray}
where $t=\left(p_\pi - p_{D^*}\right)^2$ and $u=\left(p_\pi - p_{\bar D}\right)^2$. 

\subsection{$J/\psi+\pi \rightarrow \bar D^* + D^*$}
\begin{eqnarray}
\mathcal{M}^{\mu \nu \rho}_{3a} &=& \sum_{i} F^{\nu;i}_{\pi D^* (\bar D)}(t)
\mathcal{D}_{ij}^{D}\left(p_\pi - p_{D^*}\right) F^{\mu \rho;j}_{\psi \bar D^* (D)}(t), \\ 
\mathcal{M}^{\mu \nu \rho}_{3b} &=&  \sum_{i} F^{\mu;i}_{\pi \bar D^* (
D)}(u) \mathcal{D}_{ij}^{D}\left(p_\pi - p_{\bar D^*}\right) F^{\nu \rho;j}_{\psi D^* (\bar D)}(u), \\ 
\mathcal{M}^{\mu \nu \rho}_{3c} &=& F^{\alpha \nu}_{\pi  D^* (\bar D^*)}(t) 
\mathcal{D}_{\alpha\beta}^{D^*}\left(p_\pi - p_{D^*}\right) F^{\mu \beta \rho}_{\psi \bar
D^* ( D^*)}(t), \\ 
\mathcal{M}^{\mu \nu \rho}_{3d} &=& F^{\mu \alpha}_{\pi  \bar D^* (D^*)}(u) 
\mathcal{D}_{\alpha\beta}^{D^*}\left(p_\pi - p_{\bar D^*}\right)F^{\beta \nu \rho}_{\psi 
D^* ( \bar D^*)}(u), \\ 
\mathcal{M}^{\mu \nu \rho}_{3e} &=& F^{\mu\nu\rho}_{\pi \psi \bar D^* D^*}(s,t), \\ 
 \mathcal{M}^{\mu  \nu \rho}_{3f} &=& F^{\alpha \nu}_{\pi  D^* (\bar D_1)}(t) 
\mathcal{D}_{\alpha\beta}^{D_1}\left(p_\pi - p_{D^*}\right)F^{\mu \beta \rho}_{\psi \bar
D^* ( D_1)}(t), \\ 
\mathcal{M}^{\mu \nu \rho}_{3g} &=&  F^{\mu \alpha}_{\pi  \bar D^* (D_1)}(u) 
\mathcal{D}_{\alpha\beta}^{D_1}\left(p_\pi - p_{\bar D^*}\right)F^{\beta \nu \rho}_{\psi 
D^* ( \bar D_1)}(u) 
\end{eqnarray}
where $t=\left(p_\pi - p_{D^*}\right)^2$ and $u=\left(p_\pi - p_{\bar D^*}\right)^2$.  

\subsection{$J/\psi+\rho \rightarrow \bar D + D$}
\begin{eqnarray}
\mathcal{M}^{\delta \rho}_{4a} &=& \sum_{i} F^{\delta;i}_{\rho D (\bar D)}(t)
\mathcal{D}_{ij}^{D}\left(p_\rho - p_D\right) F^{\rho;j}_{\psi \bar D (D)}(t), \\ 
\mathcal{M}^{\delta \rho}_{4b} &=& \sum_{i} F^{\delta;i}_{\rho \bar D ( D)}(u)
\mathcal{D}_{ij}^{D}\left(p_\rho - p_{\bar D}\right) F^{\rho;j}_{\psi (\bar D) D}(u), \\ 
\mathcal{M}^{\delta \rho}_{4c} &=& F^{\alpha\delta}_{\rho  D (\bar D^*)}(t) 
\mathcal{D}_{\alpha\beta}^{D^*}\left(p_\rho - p_D\right)F^{\beta\rho}_{\psi \bar D (D^*)}(t), \\ 
\mathcal{M}^{\delta \rho}_{4d} &=& F^{\alpha\delta}_{\rho  \bar D ( D^*)}(u) 
\mathcal{D}_{\alpha\beta}^{D^*}\left(p_\rho - p_{\bar D}\right)F^{\beta\rho}_{\psi D (\bar D^*)}(u), \\
\mathcal{M}^{\delta \rho}_{4e} &=&  F^{\delta \rho}_{\rho \psi \bar D D}(s,t), \\ 
\mathcal{M}^{\delta \rho}_{4f} &=&  F^{\alpha \delta}_{\rho  D (\bar D_1)}(t) 
\mathcal{D}_{\alpha\beta}^{D_1}\left(p_\rho - p_D\right)F^{\beta \rho}_{\psi \bar
D ( D_1)}(t), \\ 
\mathcal{M}^{\delta \rho}_{4g} &=&  F^{\alpha\delta}_{\rho  \bar D (D_1)}(u) 
\mathcal{D}_{\alpha\beta}^{D_1}\left(p_\rho - p_{\bar D}\right)F^{\beta \rho}_{\psi 
D ( \bar D_1)}(u)
\end{eqnarray}
where $t=\left(p_\rho - p_D\right)^2$ and $u=\left(p_\rho - p_{\bar D}\right)^2$. 

\subsection{$J/\psi+\rho \rightarrow \bar D + D^*$}
\begin{eqnarray}
\mathcal{M}^{\mu \delta \rho}_{5a} &=& \sum_{i} F^{\mu \delta;i}_{\rho D^* (\bar D)}(t)
\mathcal{D}_{ij}^{D}\left(p_\rho - p_{D^*}\right) F^{\rho;j}_{\psi \bar D (D)}(t), \\ 
\mathcal{M}^{\mu \delta \rho}_{5b} &=&  \sum_{i} F^{\delta;i}_{\rho \bar D (
D)}(u) \mathcal{D}_{ij}^{D}\left(p_\rho - p_{\bar D}\right)F^{\mu \rho;j}_{\psi D^* (\bar D)}(u), \\ 
\mathcal{M}^{\mu \delta \rho}_{5c} &=& F^{\mu \alpha \delta}_{\rho  D^* (\bar D^*)}(t) 
\mathcal{D}_{\alpha\beta}^{D^*}\left(p_\rho - p_{D^*}\right)F^{\beta \rho}_{\psi \bar
D ( D^*)}(t), \\ 
\mathcal{M}^{\mu \delta \rho}_{5d} &=& F^{\alpha \delta}_{\rho  \bar D (D^*)}(u) 
\mathcal{D}_{\alpha\beta}^{D^*}\left(p_\rho - p_{\bar D}\right)F^{\mu \beta \rho}_{\psi 
D^* ( \bar D^*)}(u), \\ 
\mathcal{M}^{\mu \delta \rho}_{5e} &=& F^{\mu\delta\rho}_{\rho \psi \bar D D^*}(s,t), \\ 
 \mathcal{M}^{\mu  \delta \rho}_{5f} &=& F^{\mu \alpha \delta}_{\rho  D^* (\bar D_1)}(t) 
\mathcal{D}_{\alpha\beta}^{D_1}\left(p_\rho - p_{D^*}\right)F^{\beta \rho}_{\psi \bar
D ( D_1)}(t), \\ 
\mathcal{M}^{\mu \delta \rho}_{5g} &=&  F^{\alpha\delta}_{\rho  \bar D (\bar D_1)}(u) 
\mathcal{D}_{\alpha\beta}^{D_1}\left(p_\rho - p_{\bar D}\right)F^{\mu \beta \rho}_{\psi 
D^* ( \bar D_1)}(u) 
\end{eqnarray}
where $t=\left(p_\rho - p_{D^*}\right)^2$ and $u=\left(p_\rho - p_{\bar D}\right)^2$. 

\subsection{$J/\psi+\rho \rightarrow \bar D^* + D^*$}
\begin{eqnarray}
\mathcal{M}^{\mu \nu \delta \rho}_{6a} &=& \sum_{i} F^{\mu \delta;i}_{\rho D^* (\bar D)}(t)
\mathcal{D}_{ij}^{D}\left(p_\rho - p_{D^*}\right) F^{\nu\rho;j}_{\psi \bar D^* (D)}(t), \\ 
\mathcal{M}^{\mu \nu \delta \rho}_{6b} &=&  \sum_{i} F^{\nu \delta;i}_{\rho \bar
D^* (D)}(u) \mathcal{D}_{ij}^{D}\left(p_\rho - p_{\bar D^*}\right) F^{\mu \rho;j}_{\psi D^* (\bar D)}(u), \\ 
\mathcal{M}^{\mu \nu \delta \rho}_{6c} &=& F^{\mu \alpha \delta}_{\rho  D^* (\bar D^*)}(t) 
\mathcal{D}_{\alpha\beta}^{D^*}\left(p_\rho - p_{D^*}\right)F^{\beta \nu \rho}_{\psi \bar
D^* ( D^*)}(t), \\ 
\mathcal{M}^{\mu \nu \delta \rho}_{6d} &=& F^{\alpha \nu \delta}_{\rho  \bar D^* (D^*)}(u) 
\mathcal{D}_{\alpha\beta}^{D^*}\left(p_\rho - p_{\bar D^*}\right)F^{\mu \beta \rho}_{\psi 
D^* ( \bar D^*)}(u), \\
\mathcal{M}^{\mu \nu \delta \rho}_{6e} &=& F^{\mu\nu\delta\rho}_{\rho \psi \bar D^* D^*}(s,t), \\ 
 \mathcal{M}^{\mu  \nu \delta \rho}_{6f} &=& F^{\mu \alpha \delta}_{\rho  D^* (\bar D_1)}(t) 
\mathcal{D}_{\alpha\beta}^{D_1}\left(p_\rho - p_{D^*}\right)F^{\beta \nu \rho}_{\psi \bar
D^* ( D_1)}(t), \\ 
\mathcal{M}^{\mu \nu \delta \rho}_{6g} &=&  F^{\alpha\nu \delta}_{\rho  \bar D^* ( D_1)}(u) 
\mathcal{D}_{\alpha\beta}^{D_1}\left(p_\rho - p_{\bar D^*}\right)F^{\mu \beta \rho}_{\psi 
D^* ( \bar D_1)}(u), \\
\mathcal{M}^{\mu \nu \delta \rho}_{6h} &=& \sum_{i} F^{\mu \delta;i}_{\rho  D^*
( \bar D^*_0)}(t)
\mathcal{D}_{ij}^{D^*_0}\left(p_\rho - p_{D^*}\right)F^{\nu \rho;j}_{\psi  \bar D^*
(D^*_0)}(t), \\
\mathcal{M}^{\mu \nu \delta \rho}_{6i} &=& \sum_{i} F^{\nu\delta;i}_{\rho \bar
D^* ( D^*_0)}(u)
\mathcal{D}_{ij}^{D^*_0}\left(p_\rho - p_{\bar D^*}\right) F^{\mu \rho;j}_{\psi  D^* (\bar D^*_0)}(u)   
\end{eqnarray}
where $t=\left(p_\rho - p_{D^*}\right)^2$ and $u=\left(p_\rho - p_{\bar D^*}\right)^2$.

\bibliography{njl.bib}

\end{document}